\documentclass[amsmath,amssymb,amsfonts,pre,reprint,bibnotes,showpacs,showkeys,longbibliography]{revtex4-1}

\usepackage{graphicx}
\usepackage{subfigure}
\usepackage{amsmath}
\usepackage{amsfonts}
\usepackage{amssymb}
\usepackage{natbib}
\usepackage{relsize}
\usepackage{sidecap}

\begin{document}


\title{Mean-field Density Functional Theory of a Three-Phase Contact Line}


\author{Chang-You Lin}
\email{changyoul@gmail.com}
\thanks{Corresponding author}
\author{Michael Widom}
\email{widom@andrew.cmu.edu}
\author{Robert F. Sekerka}
\email{sekerka@cmu.edu}
\affiliation{Department of Physics, Carnegie Mellon University, Pittsburgh, PA 15232, USA}


\date{\today}

\begin{abstract}
A three-phase contact line in a three-phase fluid system is modeled by a mean-field density functional theory. We use a variational approach to find the Euler-Lagrange equations. Analytic solutions are obtained in the two-phase regions at large distances from the contact line. We employ a triangular grid and use a successive over-relaxation method to find numerical solutions in the entire domain for the special case of equal interfacial tensions for the two-phase interfaces. We use the Kerins-Boiteux formula to obtain a line tension associated with the contact line. This line tension turns out to be negative. We associate line adsorption with the change of line tension as the governing potentials change.
\end{abstract}


\maketitle

\section{Introduction}

\begin{figure}
\includegraphics[clip,scale=0.35]{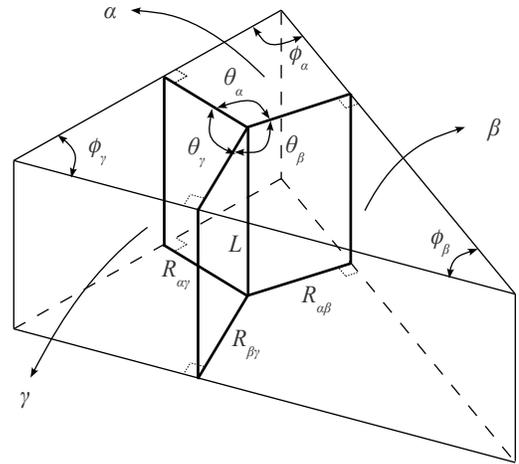}
\caption{Geometry of a system with a three-phase contact line and three interfaces within a triangular prism. The contact line is a straight line perpendicular to the base and the cap of the prism, which are Neumann triangles, and each of the interfaces is also perpendicular to the lateral boundary of the prism. The regions divided by the three interfaces contain the three bulk phases, which are labeled by $\alpha$, $\beta$, and $\gamma$. $\theta_{\alpha}$, $\theta_{\beta}$, and $\theta_{\gamma}$ are the equilibrium dihedral angles among the interfaces, and $\phi_{\alpha}$, $\phi_{\beta}$, $\phi_{\gamma}$ are the corresponding supplementary angles. $L$ is the length of the contact line, and $R_{\alpha \beta}$, $R_{\beta \gamma}$, and $R_{\gamma \alpha}$ are the distances from the contact line to the three lateral faces of the prism. We assume translational symmetry along $L$, so the problem is two-dimensional.}
\label{fig:domain}
\end{figure}

Studies of contact angle play an important role for the understanding of wetting phenomena in many systems, such as adhesives \cite{adamson1997physical}, liquid droplet spreading \cite{fukai1995wetting}, and cell adhesion \cite{sackmann2002cell}. Although contact angles can be measured, their theoretical computation can be complicated. They are affected by many factors, such as surface tension, line tension, temperature, composition of the system, and impurities, especially surfactants. Here, we focus our attention on a three-phase fluid system (Fig.~\ref{fig:domain}), where the three-phase contact line (briefly contact line) is the line where three interfaces and bulk phases meet. In this case, line tension is the excess grand potential per unit length of the contact line, which is a collective effect arising from inhomogeneities of intermolecular forces around the contact line, such as van der Waals, hydration, electrostatic, and steric forces (see \cite{adamson1997physical}). The relevant forces can be short range \cite{harkins1937linear,buff1957curved,widom1990line,szleifer1992surface,qu1999line}, or long range, the latter of which have been treated by the membrane method \cite{Derjaguin1965,Starov1980,indekeu1992line,indekeu1994line,solomentsev1999microscopic,getta1998line,bauer1999quantitative} or in terms of interacting surfaces \cite{rusanov1982surface,rusanov1996thermodynamics,rusanov1999classification,deFeijter1972transition,toshev1981some}. For a review see \cite{amirfazli2004status}. In this paper, we deal only with short range forces so the problem can be formulated in terms of local densities.

We model our system containing a three-phase contact line in the framework of general mean-field density functional theory by means of a diffuse interface model, where the imbalance of intermolecular forces is modeled by a potential function and a gradient energy of the chemical constituents. Thermodynamic-based functional theories incorporated with diffuse interfaces were first introduced by Lord Rayleigh  \cite{rayleigh1892theory}, followed by many others \cite{van1893thermodynamic,cahn1958free,bongiorno1976molecular,anderson1998diffuse}. They show good agreement with available experiments (see \cite{amirfazli2004status}). For a comprehensive introduction of density functional methods to problems involving interfaces, see Rowlinson and Widom \cite{rowlinson2002molecular}. Similar methods, known as phase-field models \cite{sekerka2001fundamentals,chen2002phase}, have been introduced to solve dynamical problems, such as moving boundary problems \cite{caginalp1986analysis,wheeler1992phase,wang1993thermodynamically}. Our model relates to a ternary solution (actually a pseudo binary) and employs a different potential from that of Widom \textit{et al.}, \cite{szleifer1992surface}, which is also a two-density model. Our potential is symmetric in the densities and is easy to relate analytically to measurable physical quantities in the far-field limit.

We consider three bulk fluid phases in a multicomponent system. As illustrated in Fig.~\ref{fig:domain}, the geometry of the system is a triangular prism. The three planer interfaces $\alpha \beta$, $\beta \gamma$ and $\gamma \alpha$ meet at a three-phase contact line of length $L$ and divide the system into three bulk phases $\alpha$, $\beta$, and $\gamma$, which subtend dihedral angles $\theta_{\alpha}$, $\theta_{\beta}$, and $\theta_{\gamma}$. Each of the interfaces is perpendicular to one of the lateral faces of the prism. The base and the cap of the prism are Neumann triangles, which are perpendicular to the contact line and the three interfaces. The distances from the contact line to the lateral boundaries of the domain are $R_{\alpha \beta}$, $R_{\beta \gamma}$, and $R_{\gamma \alpha}$. $LR_{ij}$ is the area of the interface $ij$. We treat this system in a regime where gravity is negligible. Due to the translational symmetry along $L$, the problem is effectively two-dimensional. Ultimately we consider the limit in which all $R_{ij}\rightarrow \infty$.


Classically, the problem is usually treated by regarding the interfaces to be mathematical planes (zero thickness). Since the interfacial tension is the excess grand potential per unit area, the equilibrium angles can be obtained by requiring zero variation of the excess grand potential for an infinitesimal variation of the location of the three-phase contact line. The well-known result is
\begin{equation}\label{eq:eqangle}
\frac{\sin \theta_{\alpha }}{\sigma_{\beta \gamma}} = \frac{\sin \theta_{\beta}}{\sigma_{\alpha  \gamma}} = \frac{\sin \theta_{\gamma}}{\sigma_{\alpha \beta}},
\end{equation}
where $\sigma_{\alpha \beta}$, $\sigma_{\beta \gamma}$, and $\sigma_{\alpha \gamma}$ are the interfacial tensions. In this way, the interfacial tensions can be related to a Neumann triangle, whose three side are proportional the interfacial tensions and whose three angles are the supplementary angles of the three dihedral angles. For example, in Fig.~\ref{fig:domain}, $\phi_{\beta}=\pi-\theta_{\beta}$. However, the classical model does not include the diffuse nature of the interface, nor possible complexity near the contact line.

\section{Density Functional Model}\label{sec:dfmodel}

We are interested in a thermodynamically-based description of a system which is inhomogeneous because of the interfaces and the three-phase contact line. We follow the thermodynamic methods of Gibbs, which amounts to choosing the grand canonical ensemble \cite[p~228]{gibbs1928collected} in statistical mechanics. Thus, densities of chemical components as well as entropy density are allowed to vary, while the conjugate field variables are held fixed. Assuming that the grand potential of the entire system exists, the excess grand potential $\Omega_{xs}$ due to the inhomogeneity of the system can be defined as
\begin{equation}\label{eq:omega1}
\Omega_{xs} = \Omega - \Omega_b,
\end{equation}
where $\Omega$ is the grand potential of the entire system and $\Omega_b$ is the sum of the grand potentials of the three bulk phases as if they shared the entire volume. Due to the homogeneity of the bulk phases, we have $\Omega_b = -pV$, where $p$ and $V$ are the common pressure and the total volume of the bulk phases, respectively. Thus,
\begin{equation}\label{eq:omega3}
\Omega_{xs} = \Omega + pV.
\end{equation}
By convention \cite[ch~8]{rowlinson2002molecular}, $\Omega_{xs}$ can be regarded as arising from two kinds of inhomogeneities, one associated with the contact line and the other associated with the interfaces, i.e.
\begin{equation}\label{eq:omegaxsconvention}
\Omega_{xs} = L \tau + L R_{\alpha \beta} \sigma_{\alpha \beta} + L R_{\beta \gamma} \sigma_{\beta \gamma} + L R_{\gamma \alpha} \sigma_{\gamma \alpha},
\end{equation}
where $\tau$ is the line tension, and $\sigma_{ij}$ is the interfacial tension of the interface $ij$ far from the contact line.  $L$ and $R_{ij}$ are defined in Fig.~\ref{fig:domain}. According to this convention, $\tau$ is defined as if the interfaces, with their far field values of $\sigma_{ij}$, extend all the way to the triple line where they meet. The form \eqref{eq:omegaxsconvention} of excess grand potential is to be understood in the limit of all $R_{ij}\rightarrow \infty$.

Following Rowlinson and Widom \cite{rowlinson2002molecular} we assume that $\Omega_{xs}$ can be expressed as the integral of a density $\psi(\mathbf{x})$ of the excess grand potential, so
\begin{equation}\label{eq:omegaxs}
\Omega_{xs} = L \int \psi(\mathbf{x}) \mathrm{d}A.
\end{equation}
In mean field density functional theory, $\psi(\mathbf{x})$ is assumed to be a functional of the number densities of the chemical components $\rho_i$, $i=1,2,\cdots,c$, for a $c$-component system, and $\rho_{c+1}=s$, the entropy density. Symbolically,
\begin{equation}
\psi(\mathbf{x}) =\psi\left[ \{\rho_i(\mathbf{x})\}_{i=1, \cdots ,c+1} \right],
\end{equation}
which also depends on the set of conjugate field variables $\{\mu_i\}_{i=1, \cdots ,c+1}$, where $\mu_1, \mu_2,\cdots,\mu_c$ are chemical potentials and $\mu_{c+1}=T$, the temperature. $\psi(\mathbf{x})$ is a function of densities and field variables plus a gradient energy correction, 
\begin{equation}\label{eq:psiapprox}
\psi(\mathbf{x}) = F\left( \{\rho_i(\mathbf{x})\}_{i=1, \cdots ,c+1}\right) + G \left( \{ \nabla\rho_i(\mathbf{x})\}_{i=1 \cdots c+1} \right),
\end{equation}
where $F$ is a local density of the excess grand potential, an approximation sometimes called point-thermodynamics \cite[p~43]{rowlinson2002molecular}, and $G$ is the density of gradient energy, which is usually taken to be a linear function of the $\lvert \nabla \rho_i(\mathbf{x}) \rvert^2$. The minimization of $\Omega_{xs}$ is analogous to the minimization of the integral of a Lagrangian, whose role here is played by $\psi(\mathbf{x})$. Then, the terms in $\lvert \nabla \rho_i(\mathbf{x}) \rvert^2$ play the role of kinetic energies and $F$ plays the role of the negative of the potential energy.

For a homogeneous bulk phase, there is no gradient energy and the excess grand potential density $\psi(\mathbf{x})=F=0$ where
\begin{equation}
\begin{split}
F & = \omega + p = e - T s - \sum_{i=1}^{c} \mu_i \rho_i + p \\
  & = e - \sum_{i=1}^{c+1} \mu_i \rho_i + p = 0 \mbox{ (bulk phase)}.
\end{split}
\end{equation}
Here $\omega$ is the uniform density of the grand potential; whereas, $e$, $s$, and $\rho_i$ are the densities of the internal energy, the entropy, chemical constituents that are uniform in each bulk phase. Assuming the densities of the state variables for the inhomogeneous part of the system have a similar relation to those in the bulk phases, we approximate $F$ for the entire system as
\begin{equation}
F = e - \sum_{i=1}^c \mu_i \rho_i -T s + p = e - \sum_{i=1}^{c+1} \mu_i \rho_i + p,
\end{equation}
where $e=e\left(\{\rho_i(\mathbf{x})\}_{i=1, \cdots ,c+1} \right)$ is the non-convexified internal energy as a function of the non-uniform $c+1$ densities. Since $p=p\left(\{\mu_i\}_{i=1, \cdots ,c+1} \right)$ is the common pressure of the bulk phases, it only depends on the set $\{\mu_i\}$. In general, $e \left(\{\rho_i(\mathbf{x})\}_{i=1, \cdots ,c+1} \right)$ is a non-convex function that has three potential wells and the bulk phases are given by a common tangent plane construction. Thus, $F\geq 0$ because the terms $-T s- \sum_{i=1}^c \mu_i \rho_i +p$ represent the subtraction of the common tangent plane of the bulk phases from the non-convexified internal energy. Therefore, the three potential wells that correspond to the bulk phases are located at $F=0$, where each is locally tangent to that plane. Note that $e -T s$ is the Helmholtz free energy density, as for a bulk phase. By means of an approximation discussed by \cite[p~60]{rowlinson2002molecular} and \cite{djikaev2004geometric}, we can reduce this model that depends on $c+1$ densities to an approximate model that depends on only $c$ densities, $\rho_1$, $\rho_2$, ..., and $\rho_c$. This amounts to assuming that
$\partial e/\partial s= T$, as it would for a bulk phase \cite{taylor2005adsorption}.
Thus, $\partial (e-Ts)/\partial s= 0$, so the form \eqref{eq:psiapprox} of $\psi(\mathbf{x})$ is approximated by
\begin{equation}\label{eq:psiapprox2}
\psi(\mathbf{x})=F(\{ \rho_i(\mathbf{x})\}_{i=1, \cdots ,c}) + G \left( \{ \nabla\rho_i(\mathbf{x})\}_{i=1, \cdots ,c} \right),
\end{equation}
where $G$, as a correction of $F$, is assumed to be only a function of the gradients of $c$ densities as well. $F$ also depends on the fields $\mu_i$ and $T$.

\subsection{Model for Uniform Molar Volume}

In this paper, we treat a ternary system under the constraint of constant and uniform total molar volume. We obtain a tractable problem by introducing an explicit potential that is symmetric with respect to the three chemical components.

For a ternary system, $c=3$. Under the simplifying constraint of constant total molar volume, 
\begin{equation}\label{eq:sumofc}
\rho_1 + \rho_2 + \rho_3 = \rho = \mbox{constant}.
\end{equation}
With this constraint, the system we treat is actually a pseudo-binary system that can be described by two independent concentrations, say $\rho_1$ and $\rho_2$. Moreover, this constraint means that the conjugate thermodynamic variables of $\rho_1$ and $\rho_2$ are the chemical potential differences $M_1=\mu_1-\mu_3$ and $M_2=\mu_2-\mu_3$, where the $\mu_i$ would correspond to a system with variable molar volume. In symmetric form, our potential is
\begin{equation}\label{eq:fapprox1}
F(\rho_1,\rho_2,\rho_3)= B \sum_{i=1}^3 \frac{(\rho_i-\rho_a)^2(\rho_i-\rho_b)^2}{\rho^4},
\end{equation}
where $B$ is constant with the units of energy per unit volume, and $\rho_a$ and $\rho_b$ are parameters (units of concentration), which may depend on $T$ and the $M_i$. By imposing the constraint $\rho_a+2\rho_b=\rho$, we locate the three wells at symmetric positions.

We introduce the notation $X_i \equiv \rho_i/\rho$ as the mole fraction of chemical constituent $i$ ranging from 0 to 1, $a=\rho_a/\rho$, and $b=\rho_b/\rho=(1-a)/2$. The constraint \eqref{eq:sumofc} reduces to $X_1 + X_2 + X_3 = 1$. The potential \eqref{eq:fapprox1} can be expressed as
\begin{equation}\label{eq:potential0}
\frac{F}{B}=f(X_1,X_2,X_3)=\sum_{i=1}^3 (X_i-a)^2(X_i-b)^2.
\end{equation}
The function $f(X_1,X_2,X_3)$ was originally introduced by Eldred \cite{Eldred}. In terms of independent variables, one has a two-variable function,
\begin{equation}\label{eq:potential1}
f(X_1,X_2)\equiv f(X_1,X_2,1-X_1-X_2).
\end{equation}

\begin{SCfigure}
\includegraphics[scale=0.25]{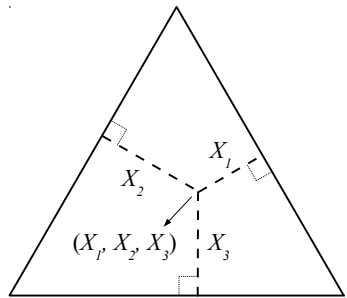}
\caption{Gibbs triangle. The summation of the distances from any point $(X_1,X_2,X_3)$ inside the triangle to the three sides of the triangle is equal to one, i.e. $X_1+X_2+X_3=1$.}
\label{fig:gibbstri}
\end{SCfigure}

\begin{figure}
\subfigure[\;a=2/3]{
\label{subfig:fpotentiala2ov3}
\includegraphics[scale=0.21]{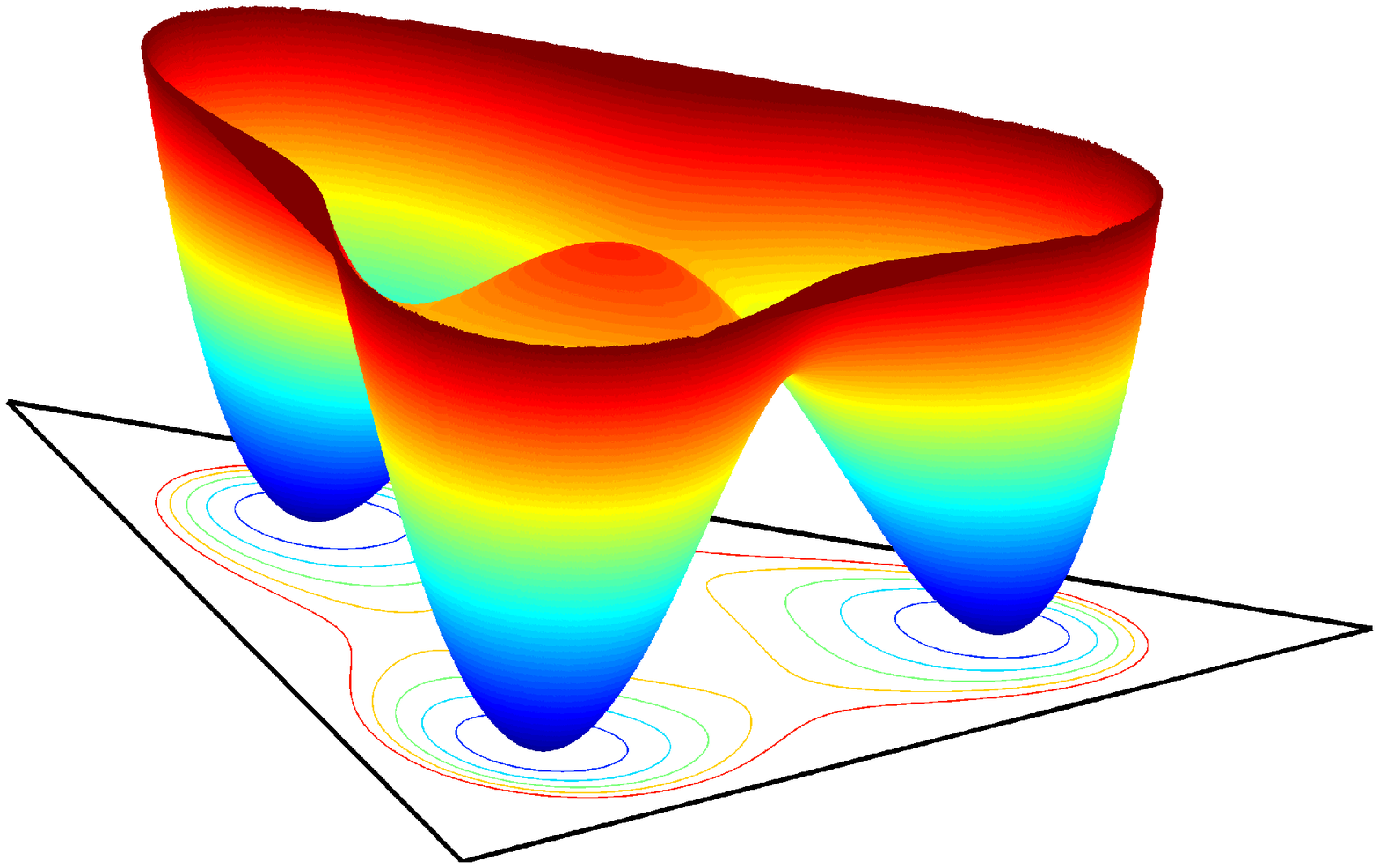}}
\subfigure[\;a=1/6]{
\label{subfig:fpotentiala1ov6}
\includegraphics[scale=0.21]{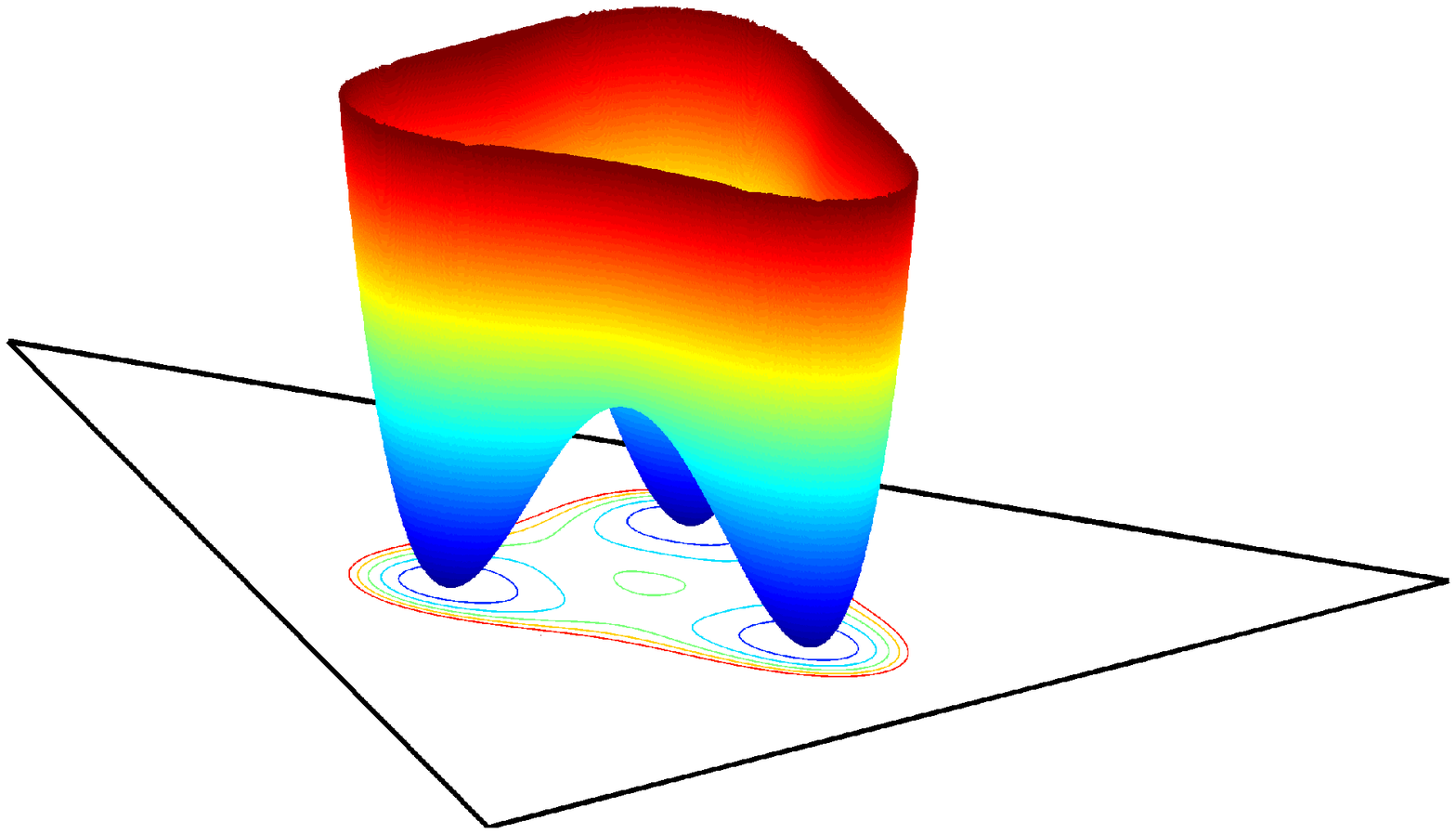}}
\caption{The three-fold symmetric potential with contours plotted on the base of the Gibbs triangle. \subref{subfig:fpotentiala2ov3} For $a = 2/3$, the three minima are located between vertices and the center. There is a local maximum at the center. \subref{subfig:fpotentiala1ov6} For $a=1/6$, the three minima are located between the mid edges and the center.}
\end{figure}

The combination of $(X_1,X_2,X_3)$ can be illustrated by a point in the Gibbs triangle as shown in Fig.~\ref{fig:gibbstri}. The compositions of the bulk phases are $(a,b,b)$, $(b,a,b)$, and $(b,b,a)$. When $a>1/3$, the three minima are located between vertices and the center of the Gibbs triangle, as illustrated in Fig.~\ref{subfig:fpotentiala2ov3}. The potential has a local maximum at $X_1=X_2=X_3=1/3$. For $a<1/3$, the three minima are rotated by $30^{\circ}$ to positions between mid edges and the center of the Gibbs triangle, as illustrated in Fig.~\ref{subfig:fpotentiala1ov6}.

%

In the reduced approximation \eqref{eq:psiapprox2} of $\psi(\mathbf{x})$, we assume that the gradient energy density is a linear function of the squares of the gradients of the mole fractions of each chemical component:
\begin{equation}\label{eq:gradient0}
\frac{G}{B} = g(\nabla X_1, \nabla X_2, \nabla X_3) = \sum\limits_{i=1}^{3} \frac{\ell_i^2}{2} \left|  \nabla X_i \right|^2,
\end{equation}
where $\ell_i$ are positive constants (with dimensions of length) associated with each chemical component. With the constraint $\nabla X_3 = - \nabla X_1 - \nabla X_2$, the form \eqref{eq:gradient0} of gradient energy density turns into a two-variable function,
\begin{equation}\label{eq:gradient1}
g(\nabla X_1, \nabla X_2)  \equiv g(\nabla X_1, \nabla X_2, - \nabla X_1 - \nabla X_2).
\end{equation}

Thus, by inserting the explicit forms of potential \eqref{eq:potential0} and gradient energy density \eqref{eq:gradient0} into the form \eqref{eq:psiapprox2} of $\psi(\mathbf{x})$, the excess grand potential in our model \eqref{eq:omegaxs} becomes
\begin{equation}\label{eq:omegaxsgeneral}
\Omega_{xs} = B L \int_A \left[ f( X_1, X_2) +  g( \nabla X_1,\nabla X_2) \right] \mathrm{d}A.
\end{equation}

\subsection{Euler-Lagrange Equations}


In equilibrium, we require $\delta \Omega_{xs} = 0$ for infinitesimal variations of $X_1$ and $X_2$. To constrain total mole number of a finite system, we could add two Lagrange multipliers for $X_1$ and $X_2$ to the integrand. However, we effectively work on an open system with infinite domain and fixed parameters $\mu_i$ and $T$, so particle conservation is not an issue and the Lagrange multipliers are effectively zero. From another point of view, the bulk phase is reached when the distance from the three-phase contact line to the boundary is large compared to the diffuse region of the contact line. This implies that the mole fractions should satisfy the boundary condition $\nabla X_i \cdot \hat{n} = 0$, where $\hat{n}$ is the unit outward normal to the physical domain. Thus, we obtain two coupled Euler-Lagrange equations
\begin{equation}\label{eq:eleq}
\begin{split}
\frac{\partial f}{\partial X_1} - (\ell_1^2+\ell_3^2) \nabla^2 X_1 - \ell_3^2 \nabla^2 X_2 &= 0, \\
\frac{\partial f}{\partial X_2} - (\ell_2^2+\ell_3^2) \nabla^2 X_2 - \ell_3^2 \nabla^2 X_1 &= 0. 
\end{split}
\end{equation}



%
%

\subsection{Asymptotic Analysis in Far-Field}

\begin{figure}
\includegraphics[clip,scale=0.23]{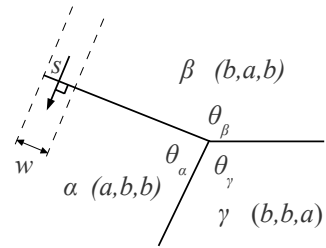}
\caption{Diagram of a transition between two bulk phases at a distance far from the three-phase contact line. In a far field limit, the transition from bulk phase $\beta$ to bulk phase $\alpha$ is represented by the transition between two wells from minimum $(b,a,b)$ to minimum $(a,b,b)$ in the Gibbs triangle. This corresponds to the transition along a one-dimensional coordinate $s$ perpendicular to the interface in spatial space. Here, $w$ is the width of an area at a distance far from the contact line.}
\label{fig:transition}
\end{figure}

In order to make a connection with the sharp interface limit of our mean-field density model, we consider a transition from phase $\beta$ to phase $\alpha$ in the far-field regime, which is far from the three-phase contact line relative to the interfacial width. This is illustrated in Fig.~\ref{fig:transition}, which corresponds to the transition from the minimum of one well to the minimum of another. Consistent with our potential function, $X_3=b$ is a constant in this region, which also satisfies the boundary condition $\nabla X_3 \cdot \hat{n}=0$. Therefore, the problem is essentially a one dimensional problem in a single variable, which we take to be $X_1$.

With $X_3=b$, we replace $X_2=1-b-X_1$, and substitute $\nabla^2 X_i= \mathrm{d}^2 X_i/\mathrm{d}s^2$ in the form of \eqref{eq:omegaxsgeneral} of excess grand potential, where $s$ is a coordinate perpendicular to the $\alpha \beta$-interface measured from $\beta$ to $\alpha$. The excess grand potential in the far field regime reduces to
\begin{equation}\label{eq:omegaxs1d}
\Omega_{xs} = BLw \mathlarger{\int} \left[ H(X_1) + \frac{\alpha_{12}}{2} \left( \frac{\mathrm{d}X_1}{\mathrm{d}s} \right)^2 \right] \mathrm{d}s,
\end{equation}
where $w$ is the width of an area in the far field regime as indicated in Fig.~\ref{fig:transition}, $H(X_1)\equiv 2(X_1-a)^2(X_1-b)^2$, and $\alpha_{12} \equiv \ell_1^2+\ell_2^2$. The limits of integration are effectively from $-\infty$ to $\infty$.


In equilibrium, we require $\delta \Omega_{xs} = 0$ for an infinitesimal variation of $X_1$ and obtain the Euler-Lagrange equation in the far-field limit:
\begin{equation}\label{eq:ELfar}
H'(X_1) = \alpha_{12} \frac{\mathrm{d}^2 X_1}{\mathrm{d}s^2}.
\end{equation}
Then, we multiply by $\mathrm{d}X_1/\mathrm{d}s$ and integrate to obtain
\begin{equation}\label{eq:xequation}
H(X_1)=\frac{\alpha_{12}}{2}\left(\frac{\mathrm{d}X_1}{\mathrm{d}s}\right)^2,
\end{equation}
where the integration constant is zero because $H(a)=H(b)=0$ and the slope $\mathrm{d}X_1/\mathrm{d}s$ is zero for $X_1=a$ and $X_1=b$.

By solving \eqref{eq:xequation}, we obtain the far-field solution for $X_1$ at the $\alpha \beta$-interface,
\begin{equation} \label{eq:x1int12}
X_1(s)=\frac{a+b}{2} + \frac{a-b}{2} \tanh \left[\frac{s}{\delta_{int,\alpha \beta}} \right],
\end{equation}
where we choose $s=0$ as $X_1=(a+b)/2$ and define the interfacial width parameter of the $\alpha \beta$-interface as
\begin{equation}\label{eq:intwidth12}
\delta_{int,\alpha \beta} \equiv \frac{\sqrt{\alpha_{12}}}{\lvert a-b \rvert}=\frac{\sqrt{\ell_1^2+\ell_2^2}}{\lvert a-b \rvert }.
\end{equation}

Of course, $X_2=1-b-X_1$. These analytic solutions were originally found by Eldred \cite{Eldred}. Analytical far-field solutions for density profiles and interfacial tensions for a symmetric three-phase contact line but a different potential were obtained by Szleifer and Widom \cite{szleifer1992surface}.
Note that when $a=b=1/3$, the interfacial widths diverge. The contact line and the three interfaces vanish. In this case, the three chemical constituents have mole fractions of $1/3$ distributed uniformly over the entire system.

\begin{figure}
\includegraphics[scale=0.89]{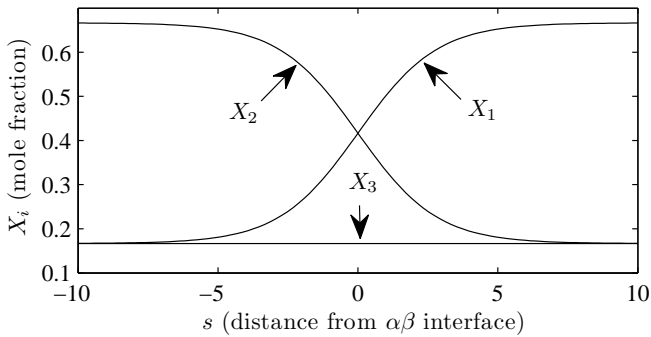}
\caption{Asymptotic far-field solutions for the mole fractions $X_I$ at the $\alpha \beta$ interfaces, for $a=2/3$. $s$ is the distance from $\beta$ to $\alpha$ perpendicular to the $\alpha \beta$ interface}
\label{fig:plotxfar}
\end{figure}

As illustrated in Fig.~\ref{fig:plotxfar}, when $s$ is negative infinity, we have $X_1=b$ and $X_2=a$, which indicates the $\beta$ bulk phase. In contrast, when $s$ is positive infinity, we obtain $X_1=a$ and $X_2=b$, which refers to the $\alpha$ bulk phase. Similarly, we can apply the same analysis for the other two interfaces and obtain solutions for $X_1$, $X_2$, and $X_3$ in the far-field limit. 

The definition of interfacial tension is the excess grand potential per unit area of interface. Thus, by inserting the relation \eqref{eq:xequation} that connects potential and gradient density into the excess grand potential \eqref{eq:omegaxs1d}, the interfacial tension of the $\alpha \beta$-interface can be expressed as
\begin{equation}\label{eq:gamma12int}
\begin{split}
\sigma_{\alpha \beta} & = \frac{\Omega_{xs}}{Lw} = B \mathlarger{\int}_{-\infty}^{\infty} \left[ H(X_1) + \frac{\alpha_{12}}{2} \left(\frac{\mathrm{d}X_1}{\mathrm{d}s}\right)^2 \right] \mathrm{d}s \\
& = B \alpha_{12} \mathlarger{\int}_{-\infty}^{\infty} \left(\frac{\mathrm{d}X_1}{\mathrm{d}s}\right)^2 \mathrm{d}s \\
& = B \sqrt{\alpha_{12}} \mathlarger{\int}_{b}^{a} \sqrt{2 H(X_1)} \mathrm{d}X_1.
\end{split}
\end{equation}
After integration, the interfacial tension of the $\alpha \beta$-interface is found to be
\begin{equation}\label{eq:gamma12}
\sigma_{\alpha \beta} = \frac{\lvert a-b \rvert^3}{3} B \sqrt{\alpha_{12}} = \frac{\lvert 3a-1\rvert^3}{24} B \sqrt{\ell_1^2 + \ell_2^2}.
\end{equation}
The interfacial tensions of the $\beta \gamma $-interface and the $\gamma \alpha$-interface can be obtained similarly.
Consistent with the classical relation of equilibrium angles \eqref{eq:eqangle}, the equilibrium angles in our model obey
\begin{equation}
\frac{\sin \theta_{\alpha}}{\sqrt{\ell_2^2 + \ell_3^2}} = \frac{\sin \theta_{\beta}}{\sqrt{\ell_1^2 + \ell_3^2}} = \frac{\sin \theta_{\gamma}}{\sqrt{\ell_1^2 + \ell_2^2}}.
\end{equation}

\section{Numerical Analysis for Symmetric Three-Phase Contact Line}\label{sec:numerical}


Due to the nonlinearity of the Euler-Lagrange equations \eqref{eq:eleq}, one cannot obtain an analytic solution for the entire domain containing the three-phase contact line. We consider a simplified symmetric contact line centered in an equilateral triangular prism.

Let $\ell_1=\ell_2=\ell_3 \equiv \ell$, where $\ell$ is a characteristic length. For convenience, we define the dimensionless coordinate $r' \equiv r/\ell$, and $\nabla'^2 \equiv \ell^2 \nabla^2$. Thus, the dimensionless form of the excess grand potential \eqref{eq:omegaxsgeneral} is given by
\begin{equation}\label{eq:dimensionlessOmega}
\Omega_{xs}'\equiv \frac{\Omega_{xs}}{BL\ell^2} = \int_A \left[f(u,v)+\bar{g}(\nabla'u,\nabla'v) \right] \mathrm{d}A',
\end{equation}
where $A' \equiv A/\ell^2$, and, for convenience of writing, we define $u=X_1$ and $v=X_2$; then $f(u,v)$ is the potential \eqref{eq:potential1}, and $\bar{g}(\nabla'u,\nabla'v)$ is the symmetric version of the gradient energy density \eqref{eq:gradient1}, but in the form
\begin{equation}
\bar{g}(\nabla'u,\nabla'v) \equiv \left| \nabla' u \right|^2 +  \left| \nabla' v \right|^2 + \nabla' u \cdot \nabla' v.
\end{equation}
Similarly, the Euler-Lagrange equations \eqref{eq:eleq} can be diagonalized and take the dimensionless form
\begin{equation}\label{eq:neweleq}
\begin{split}
\nabla'^2 X_1 - \frac{2}{3} \frac{\partial f}{\partial X_1} + \frac{1}{3} \frac{\partial f}{\partial X_2} & = 0, \\
\nabla'^2 X_2 + \frac{1}{3} \frac{\partial f}{\partial X_1} - \frac{2}{3} \frac{\partial f}{\partial X_2} & = 0.
\end{split}
\end{equation}
The dimensionless interfacial width parameter is
\begin{equation}\label{eq:deltaint}
\delta_{int}' =\sqrt{2}/\lvert a-b \rvert.
\end{equation}

\begin{figure}
\includegraphics[clip,scale=0.3]{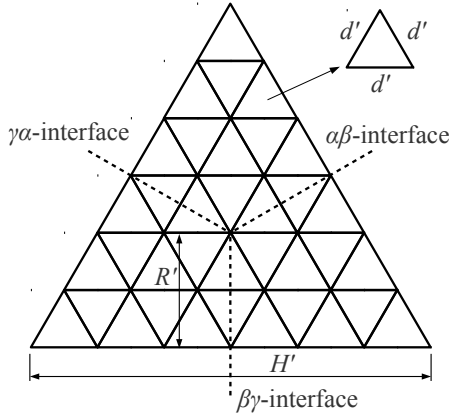}
\caption{A small triangular grid. The triangular domain is filled out by small triangles with dimensionless edge length $d'$. For convenience, each edge of the outer triangular domain is perpendicular to an interface. Also, by specific choice, the geometric center and the interfacial centers of every edge are on grid points. The triangular grid follows the rule that the number of grid points on each edge is $N=6m+1$, where $m$ is an integer. Here we take $m=1$ for illustrative purposes only.}
\label{fig:trigrids}
\end{figure}

Because of the three-fold symmetry of our system for a symmetric three-phase contact line, and the fact that the Laplacian operator is well-behaved on a triangular grid, we employ an equilateral triangular grid to resolve our special geometry. The computational domain is chosen as an equilateral triangle with physical dimension large compared to the dimensionless interfacial width \eqref{eq:deltaint}, and the grid points are determined by filling out smaller triangles with non-dimensional length $d'$ as shown in Fig.~\ref{fig:trigrids}. There are $N$ grid points on each domain edge, which is of length $H'=(N-1)d'$ and perpendicular to one of the interfaces. The distance from the contact line to each of the outer edges is $R'=(N-1)d'/(2\sqrt{3})$. For convenience, we make a special choice of grid points to allow the grid points to lie at important points of our system, such as the center of the contact line and the transition points of the far-field interfaces. To do this, the number of grid points on each edge is chosen to be $N=6m+1$, where $m$ is an integer. Therefore, the dimensionless size of each edge of the outer triangle is $H'=6md'$ and the dimensionless distance from the contact line to each edge of the outer triangle is $R'=\sqrt{3}md'$.

The asymptotic far-field solution approaches an effectively one-dimensional two-phase problem. The interfacial width is small compared to the distance from the three-phase contact line. This setup makes the far-field solutions easy to apply at the boundary of the domain. Also, the corner regions of the large triangular domain approache the bulk phases, where our potential vanishes.

\subsection{Consistent Discretization}

\begin{figure}
\subfigure[]{\label{subfig:smalltriangle}\includegraphics[clip,scale=0.8]{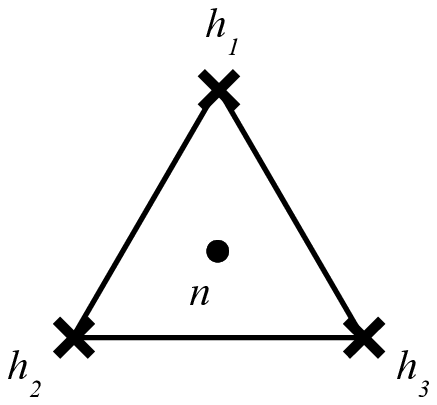}}
\subfigure[]{\label{subfig:neighbor}\includegraphics[clip,scale=0.6]{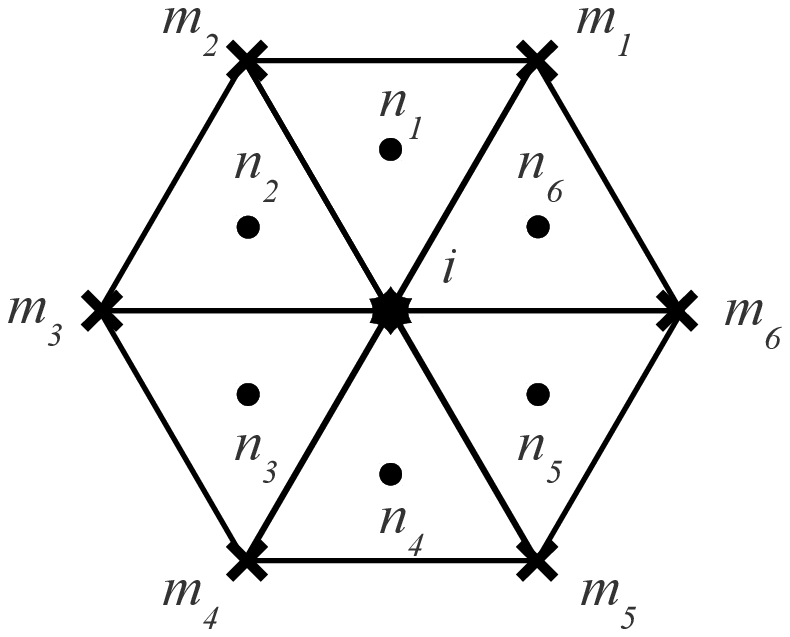}}
\caption{\subref{subfig:smalltriangle} A small triangle in the grid of the physical domain. Each equilateral small triangle has edge length $d'$ and is denoted by its center point $n$. $V^{(n)} \equiv \{h_1,h_2,h_3\}$ is the set of vertices of the small triangle denoted by $n$. \subref{subfig:neighbor} The nearest neighbors and the nearest center points for a site $i$ in a triangular grid. $NCP^{(i)} \equiv \{n_1,n_2,n_3,n_4,n_5,n_6\}$ is the set of the nearest center points for the site $i$, and $NN^{(i)} \equiv \{m_1,m_2,m_3,m_4,m_5,m_6\}$ is the set of nearest neighbors for the site $i$.}
\end{figure}

To discretize the Euler-Lagrange equations \eqref{eq:neweleq}, for a symmetric three-phase contact line based on the triangular grid in Fig.~\ref{fig:trigrids}, we employ a variation of the discrete form of the excess grand potential to avoid inconsistent discretization of the potential of $f$ and the gradient energy $g$. We approximate $u$ and $v$ as planer functions in the region of each small triangle of the triangular grid. Then, the value of $u$ and $v$ at the central point $n$ is defined as $u_n \equiv \sum_{m \in V^{(n)}} u_m/3$ and $\sum_{m \in V^{(n)}} v_m/3$, where $V^{(n)}$ is the set of vertices of the small triangle denoted by its center point $n$ (see Fig.~\ref{subfig:smalltriangle}). The discrete form of the dimensionless excess grand potential \eqref{eq:omegaxsgeneral} is approximated by evaluating the integrand at the central point of each small triangle, 
\begin{equation}\label{eq:discreteOmegaxs}
\Omega_{xs}' \sim \sum_{n \in CP} (f_n+\bar{g}_n) \Delta
\end{equation}
where $CP$ is the set of the central points of the small triangles over the entire triangular grid in Fig.~\ref{fig:trigrids}; $\Delta$ is the area of each small triangle; $f_n \equiv f(u_n,v_n)$ and $\bar{g}_n \equiv \bar{g}((\nabla'u)_n,(\nabla'v)_n)$. Since $u$ and $v$ are approximated by planer functions, we obtain 
\begin{equation}\label{eq:glinearapproxi}
\begin{split}
\bar{g}_n  \simeq & \frac{2}{3d'^2}  \sum_{(j,k) \in PV^{(n)}, j\ne k} \left[ (u_j-u_k)^2 \right. \\
& + \left. (v_j-v_k)^2 + (u_j-u_k)(v_j-v_k)\right].
\end{split}
\end{equation}
where $PV^{(n)}$ is the set of pairs of the vertices $V^{(n)}$ (Fig.~\ref{subfig:smalltriangle}) of the small triangle $n$ with edge $d'$.


At equilibrium, we require $\delta \Omega_{xs}'=0$ for the discrete form \eqref{eq:discreteOmegaxs} of the excess grand potential. From the chain rule, this is equivalent to the vanishing of the sum of the variations of all of the unknowns $(u_i,v_i)$ for each internal site $i$ of the triangular grid. According to the approximations \eqref{eq:glinearapproxi} of the gradient energy density and requiring the coefficients of $\delta u_i$ and $\delta v_i$ to vanish, we obtain the discrete Euler-Lagrange equations for each site $i$ (each internal grid point of the triangular grid),
\begin{equation}\label{eq:consistentdiscreteeleq}
\begin{split}
(\nabla'^2 u)_i - \frac{2}{3} \left(\overline{\frac{\partial f}{\partial u}}\right)_{NCP^{(i)}} + \frac{1}{3} \left(\overline{\frac{\partial f}{\partial v}}\right)_{NCP^{(i)}} = 0,\\
(\nabla'^2 v)_i + \frac{1}{3} \left(\overline{\frac{\partial f}{\partial u}}\right)_{NCP^{(i)}} - \frac{2}{3} \left(\overline{\frac{\partial f}{\partial v}}\right)_{NCP^{(i)}}  = 0,
\end{split}
\end{equation}
where $\overline{\left(\partial f/\partial u \right)}_{NCP^{(i)}} \equiv  \sum_{n \in NCP^{(i)}} (\partial f/\partial u)_n/6$ and $\overline{\left(\partial f/\partial v \right)}_{NCP^{(i)}} \equiv  \sum_{n \in NCP^{(i)}} (\partial f/\partial v)_n/6$ are the averages of $\partial f/\partial u$ and $\partial f/\partial v$ over the six nearest center points of each site $i$, $NCP^{(i)}$ (Fig.~\ref{subfig:neighbor}).
The approximate Laplacian operators according to second order Taylor's series expansions are
$(\nabla'^2 u)_i \equiv 4 (\overline{u}_{NN^{(i)}} - u_i)/d'^2$ and
$(\nabla'^2 v)_i  \equiv 4 (\overline{v}_{NN^{(i)}} - v_i)/d'^2$,
where $\overline{u}_{NN^{(i)}} \equiv \sum_{m \in NN^{(i)}} u_m /6$ and  $\overline{v}_{NN^{(i)}} \equiv \sum_{m \in NN^{(i)}} v_m /6$ are the averages of $u$ and $v$ over the six nearest neighbors of each site $i$, $NN^{(i)}$ (Fig.~\ref{subfig:neighbor}).

Note that the discrete Euler-Lagrange equations \eqref{eq:consistentdiscreteeleq} are similar to the analytic form \eqref{eq:neweleq}, except $\partial f/\partial X_1$ and $\partial f/\partial X_2$ are replaced by the average values over the six nearest central points.
After we apply the asymptotic far-field solutions as the boundary conditions of the system of algebraic equations for the triangular grid, there are $(N-2)(N-3)$ algebraic equations for the whole domain.

\subsection{Successive Over-relaxation Method}

To solve the system of coupled algebraic equations, we apply the method of successive over-relaxation (SOR) \cite{young1950iterative,frankel1950convergence}. We define error equations for the diagonalized form of the discrete Euler-Lagrange equations \eqref{eq:consistentdiscreteeleq}:
\begin{equation}\label{eq:algebraicerror}
\begin{split}
(r_u)_i & = (\nabla'^2 u)_i -\frac{\lambda}{3} \left[ 2 \left(\overline{\frac{\partial f}{\partial u}}\right)_{NCP^{(i)}} - \left(\overline{\frac{\partial f}{\partial v}}\right)_{NCP^{(i)}} \right],\\
(r_v)_i & = (\nabla'^2 v)_i -\frac{\lambda}{3} \left[- \left(\overline{\frac{\partial f}{\partial u}}\right)_{NCP^{(i)}} + 2 \left(\overline{\frac{\partial f}{\partial v}}\right)_{NCP^{(i)}}\right],
\end{split}
\end{equation}
where $ 0 \leq \lambda \leq 1$ is an adjustable parameter used to implement our numerical technique, and $(r_u)_i$ and $(r_v)_i$ are the residues that we try to make as small as practical.

In the form \eqref{eq:algebraicerror} of error equations, $\overline{\left(\partial f/\partial u \right)}_{NCP^{(i)}}$ and $\overline{\left(\partial f/\partial v \right)}_{NCP^{(i)}}$ are polynomials of the mole fractions. To avoid the complexity of numerical calculation due to the nonlinearity of these terms, at the beginning, $\lambda$ is set to be zero. After solving this simplest version of the equation by SOR, we apply that solution as the initial values of SOR for new equations in which $\lambda$ is increased by a small fraction of 1, and solve the equations again. Then, we gradually enlarge $\lambda$ and repeat this procedure until $\lambda$ reaches one.

In the updating process of SOR, we first input guessed numbers of $u_i$ and $v_i$ as initial values into error equations \eqref{eq:algebraicerror} for every site in the grid. Then, we update $u_i$ and $v_i$ for each site by
\begin{equation}
\begin{split}
u_i^{new} & =u_i^{old}- q \frac{d'^2}{4} (r_u)_i,\\
v_i^{new} & =v_i^{old}- q \frac{d'^2}{4} (r_v)_i, 
\end{split}
\end{equation}
where $q=1.86$ \cite{Taasan}. We repeat this procedure until the values of $u_i$ and $v_i$ at every site converge. 

To check convergence, we study the norm of errors after every iteration. The norms are defined as
\begin{equation}
\left\Vert r_u \right\Vert = \sqrt{\frac{\sum_{i}^{N_{tot}} (r_u)_i}{N_{tot}}} \mbox{ and } \left\Vert r_v \right\Vert = \sqrt{\frac{\sum_{i}^{N_{tot}} (r_v)_i}{N_{tot}}}.
\end{equation}
Then the convergence criteria can be defined as $\left\Vert r_u \right\Vert$ and $\left\Vert r_v \right\Vert$ are simultaneously smaller than $\epsilon$, where $\epsilon$ is a small number. Alternatively, this means $\left\Vert u_i^{new} - u_i^{old} \right\Vert$ and $\left\Vert v_i^{new} - v_i^{old} \right\Vert$ are simultaneously smaller than $q d'^2\epsilon/4$.

\subsection{Contours and Profiles}

\begin{figure}
\subfigure[\;Contour plot of the numerical solution of $X_1$]{
\label{subfig:fineplotofX1}
\includegraphics[scale=0.9]{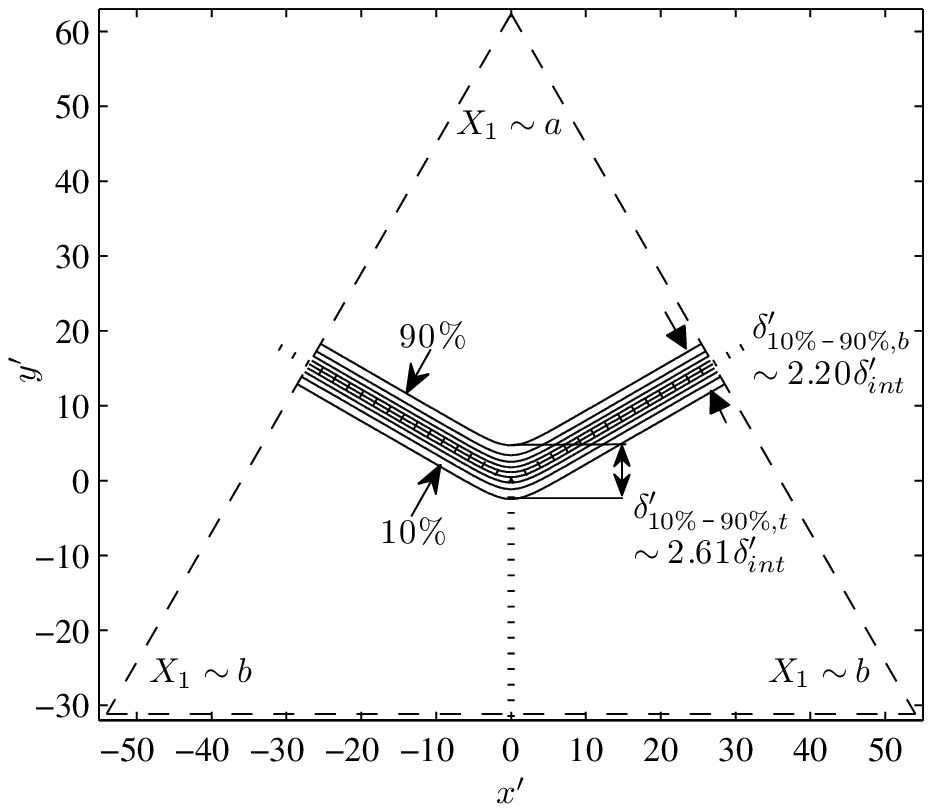}}
\subfigure[\;Profiles of $X_1$ along $x'=0$]{
\label{subfig:X1profileatx0}
\includegraphics[scale=0.88]{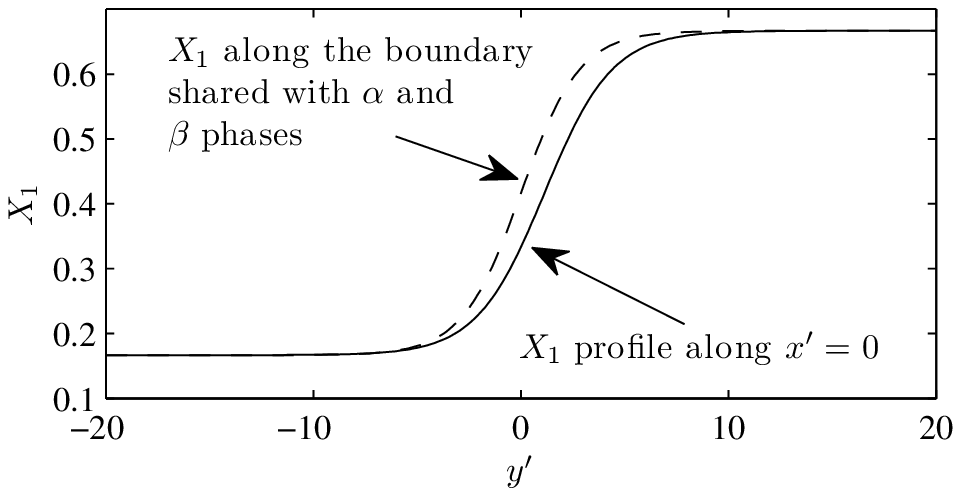}}
\caption{\subref{subfig:fineplotofX1} Contour plot of the numerical solution of $X_1$ for $a=2/3$, $m=180$, and $d'=0.1$. Here, $N=1081$ and $b=1/6$. $x'\equiv x/\ell$ and $y'\equiv y/\ell$ are the dimensionless Cartesian coordinates. The domain edge is $H' = 108$, which is large compared to the interfacial width parameter $\delta_{int}' =2\sqrt{2}$. The contours are evenly spaced from $10\%$ to $90\%$ of $ a-b $. Note that the interfacial width defined by the difference between $10\%$ and $90\%$ at the boundary,  $\delta_{10\%-90\%,b}'$, is around $2.20 \delta_{int}'$, whereas the interfacial width at the three-phase contact line, $\delta_{10\%-90\%,t}'$, is around $2.61 \delta_{int}'$. \subref{subfig:X1profileatx0} Profiles of $X_1$ along $x'=0$ (along the central vertical line of Fig.~\ref{subfig:fineplotofX1}) and the boundary shared with $\alpha$ and $\beta$ phases of the numerical solution of $X_1$ for $a=2/3$, $m=180$, and $d'=0.1$. The diffuse region of the profile along $x'=0$ is slightly widened and shifted compared to the profile along the boundary shared with $\alpha$ and $\beta$ phases.}
\end{figure}



Here, we present a numerical solution obtained from SOR. The numerical input is $a=2/3$, $d'=0.1$, and $m=180$. The error tolerance, $\epsilon$, is $10^{-8}$. Then the domain edge $H'=108$ is large compared to the interfacial width parameter $\delta_{int}' = 2\sqrt{2}$. Solutions for $X_2$ and $X_3$ are just the rotation of the solution of $X_1$ by $120^{\circ}$ and $240^{\circ}$. The solution for $X_1$ in Fig.~\ref{subfig:fineplotofX1} reveals the nature of diffuse interfaces. There is bending and slight widening of the diffuse region for $X_1$ near the three-phase contact line, which is quantified by the interfacial width defined from $10\%$ to $90\%$ isoconcentration lines. The width at the boundary, $\delta'_{10 \%-90 \%,b}$, is about $6.21 \approx 2.20 \delta_{int}'$. For comparison, the width at the contact line, $\delta'_{10 \%-90 \%,t} \approx 7.37 \approx 2.61 \delta_{int}'$, is about $20\%$ larger. $\delta'_{10 \%-90 \%,t}$ is small compared to the distance from the outer domain boundary to the contact line along any interface, which is $R' \approx 31.2 \approx 11.0 \delta_{int}'$. Also, the profile at the contact line shifts its center compared to the one at the boundary, as shown in Fig.~\ref{subfig:X1profileatx0}. Close to the boundary, the nearly parallel isoconcentration lines along the interfaces show that our domain size is close to the asymptotic regime, consistent with our intended boundary condition.

\section{Line Tension}

\subsection{Density Functional Model for Line Tension}\label{sec:dfmlinetension}


The numerical results in Sec.~\ref{sec:numerical} reveal the fact that the actual interfacial width increases slightly while approaching the three-phase contact line. This result is different than that which would be obtained by extrapolation of the far-field solution. In this section, we study the line tension which is the excess energy per unit length associated with the three-phase contact line. By convention, the line tension is defined in the form \eqref{eq:omegaxsconvention} of the excess grand potential. For a symmetric contact line, we let $R \equiv R_{\alpha \beta} = R_{\beta \gamma} = R_{\gamma \alpha}$, and $\sigma \equiv \sigma_{\alpha \beta} = \sigma_{\beta \gamma} = \sigma_{\gamma \alpha}$. In terms of the dimensionless grand potential \eqref{eq:dimensionlessOmega} with $R'\equiv R/\ell$, the dimensionless line tension is given by
\begin{equation}\label{eq:taudefsym}
\tau'\equiv \frac{\tau}{B \ell^2} = \Omega'_{xs}- 3 R' \sigma',
\end{equation}
where $\sigma'$ is the dimensionless form of the interfacial tension in the far-field limit \eqref{eq:gamma12} for a symmetric contact line, i.e.
\begin{equation}\label{eq:symsigma}
\sigma' \equiv  \frac{\sigma}{B \ell}=\frac{\sqrt{2}}{3}\lvert a-b \rvert^3 = \frac{1}{\sqrt{2}}\left(\frac{3}{2}\right)^2 \left| a-\frac{1}{3} \right|^3.
\end{equation}
From the symmetry of $\psi(\mathbf{x})$ for a symmetric contact line, the solutions for the three mole fractions have the property $X_1(r,\theta)=X_2(r,\theta-\frac{2\pi}{3})=X_3(r,\theta+\frac{2\pi}{3})$, which means each of them are given by only a rotation of $2 \pi/3$ or $-2 \pi/3$ from the others. Thus, the integration in the form \eqref{eq:omegaxsgeneral} of excess grand potential can be divided into three equal parts. By applying the boundary condition $\nabla' X \cdot \hat{n}=0$, we find that $\int \frac{1}{2} \left| \nabla' X_1 \right|^2 \mathrm{d}A'= - \int \frac{1}{2} X_1 \nabla'^2 X_1 \mathrm{d}A'$. Thus, the dimensionless line tension is given by
\begin{equation}\label{eq:core}
\begin{split}
\tau' = & 3 \mathlarger{\int}_A \left[ (X_1-a)^2 (X_1-b)^2 - \frac{1}{2} X_1 \nabla'^2 X_1 \right] \mathrm{d}A' \\
& - \sqrt{2}\lvert a-b \rvert^3 R'.
\end{split}
\end{equation}

We find, however, that evaluation of the form \eqref{eq:core} of dimensionless line tension is sensitive to the choice of boundary condition, which may result from the inconsistency between the numerical evaluation of the excess grand potential and the analytic interfacial tension. Instead, we use a formula for line tension derived by Kerins and Boiteux \cite{KBpaper}, which transfers the second term of the form \eqref{eq:core} of line tension into a surface integral and combines it with the first term. In this integral form, the integrands will vanish at distances far from the three-phase contact line, which means it is insensitive to domain size for a sufficiently large domain. According to the Kerins-Boiteux formula, the dimensionless line tension in given by 
\begin{equation}\label{eq:dimensionlessKB}
\tau' = \int_A \left[-f(u,v)+g(\nabla' u,\nabla' v) \right] \mathrm{d}A'.
\end{equation}

Numerically, we can discretize the integral in Eq.~\eqref{eq:dimensionlessKB} by employing a triangular grid as in Fig.~\ref{fig:trigrids}, so
\begin{equation}\label{eq:discretetau}
\tau' \sim \sum_{n \in CP} \left[-f_n + \bar{g}_n\right] \Delta,
\end{equation}
where $f_n$ and $\bar{g}_n$ are defined in the discrete form \eqref{eq:discreteOmegaxs} of the excess grand potential and the approximation \eqref{eq:glinearapproxi} of the gradient energy density. Then we can utilize the numerical method developed in Sec. \ref{sec:numerical} to obtain the numerical value of dimensionless line tension in our model.

\subsection{Evaluation of Line Tension}\label{sec:evaluationoflinetension}



\begin{figure}
\includegraphics[scale=0.91]{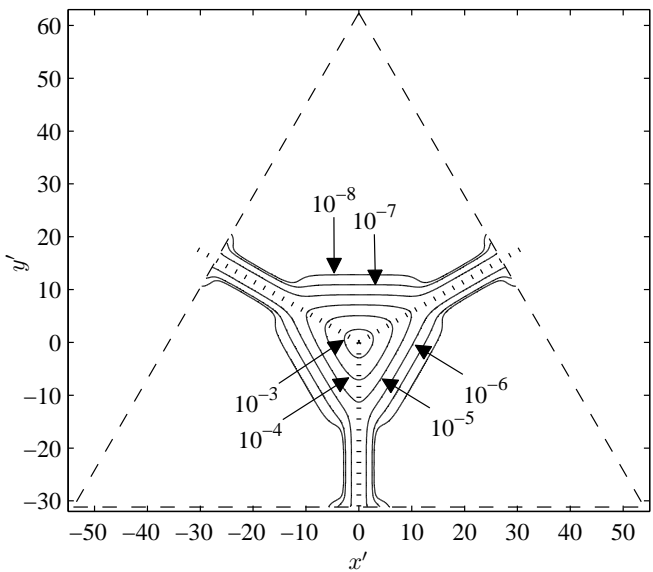}
\caption{A contour plot on a logarithmic scale at the integrand of the Kerins-Boiteux formula over a triangular domain with a threshold of $10^{-8}$. The parameters for this calculation are $m=180$, $d'=0.1$, and $a=2/3$. The major contribution of the integrand with values from $10^{-2}$ to $10^{-5}$ is confined in a core region with a dimension of $2$ to $3$ interfacial widths, $\delta_{int}'$, near the three-phase contact line. The minor contribution, which has values ranging from $10^{-5}$ to $10^{-8}$, is outside the core region and along the three interfaces with width of about $1.5 \delta_{int}'$. This shows that the integrands within the bulk phases are significantly smaller compared to the core region and along the three interfaces. Also, the nonzero contours along the three interfaces are nearly parallel near the outer boundary.}
\label{fig:KBintegrand}
\end{figure}

We perform a numerical evaluation of the integrand of the discrete form \eqref{eq:discretetau} of the Kerins-Boiteux formula. Fig.~\ref{fig:KBintegrand} shows a contour plot of the integrand on a logarithmic scale. A similar plot on a normal scale can be found in Taylor and Widom \cite{taylor2005adsorption}. The integrand decays exponentially for the most of the domain. The major contribution of the integrand is approximately within the range from $10^{-2}$ to $10^{-5}$ at a core region centered at the three-phase contact line with dimension of $2$ to $3$ times $\delta_{int}'$. The minor contribution, which is considered to be from $10^{-5}$ to $10^{-8}$, is distributed outside the core region and along the three interfaces with a width of about $1.5 \delta_{int}'$. The integrand in the rest of the domain is less than $10^{-8} $ and is essentially negligible compared to the one close to contact line and along the three interfaces. Theoretically, the potential and gradient energy density are zero within bulk phases and $-f+g \rightarrow 0$ in the interfaces far from the core, so the small but non-vanishing values of $-f+g$ along the interfaces results from numerical errors. Also, we can see that the contours of the integrand begin to bend at the far-field boundary of the domain, which may relate to the errors associated with applying the far-field solution as the boundary condition for a finite domain. Note that the contours along the three interfaces are nearly parallel except close to the contact line and the boundary. This suggests that the numerical evaluation of the Kerins-Boiteux integral over these areas leads to an error that is approximately proportional to $R'$. Our numerical results also show that when $d'$ is smaller, the distribution of the integrand within the core region is sharper, with a slightly larger maximum value, and decays faster, which means that the integrands along the three interfaces and the boundary decrease when $d'$ becomes smaller. So, we assume that the numerical error of the evaluation of the Kerins-Boiteux formula is proportional to $R'$ and depends on $d'$.

\begin{figure}
\includegraphics[scale=0.9]{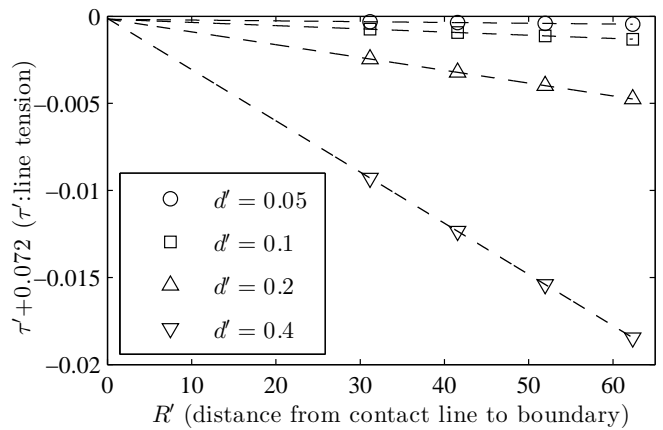}
\caption{Line tension $\tau'$ as a function of $R'$ for various $d'$ values calculated by Kerins-Boiteux formula with $a=2/3$. $\tau'$ is calculated for $d'=0.05$, $0.1$, $0.2$, and $0.4$ and  $R' \approx 31.2$, $41.2$, $52.0$, and $62.4$. The dashed lines are the linear extrapolations of the line tensions from various $R'$ toward $R'=0$ for each $d'$.}
\label{fig:extauKBvsR}
\end{figure}

\begin{figure}
\includegraphics[scale=0.9]{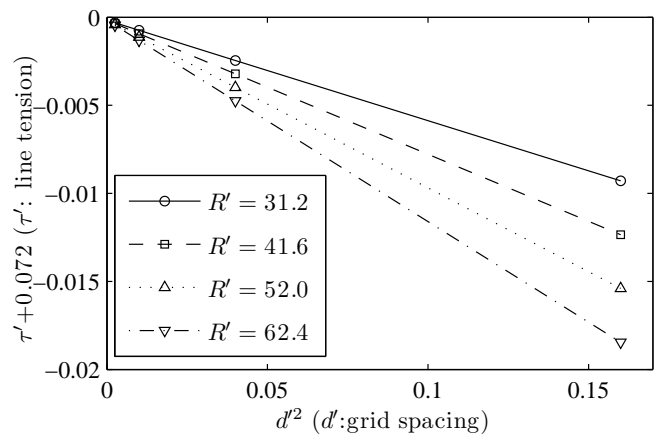}
\caption{Line tension as a function of $d'^2$ at various $R'$ values calculated by the Kerins-Boiteux formula with $a=2/3$. $\tau'$ is calculated for $d'=0.05$, $0.1$, $0.2$, and $0.4$ and  $R' \approx 31.2$, $41.2$, $52.0$, and $62.4$.}
\label{fig:tauKBvsdsq}
\end{figure}

To test this, we use the Kerins-Boiteux formula to calculate values of line tension, $\tau'$. As shown in Fig.~\ref{fig:extauKBvsR}, the $\tau'$ value is nearly proportional to $R'$ for each $d'$. We take grid spacings, $d'=0.05$, $0.1$, $0.2$, and $0.4$, and domain sizes, $R' \approx 31.2$, $41.2$, $52.0$, and $62.4$, which are relatively large compared to the size of the three-phase contact line, roughly $7.37 \approx 2.61 \delta_{int}'$. By linear extrapolation from the results in Fig.~\ref{fig:extauKBvsR}, we find that the values of $\tau'$ for different $d'$ roughly meet at $R'=0$. Moreover, from Fig.~\ref{fig:tauKBvsdsq}, we find that the dominant numerical error of $\tau'$ comes from a $d'^2$ term for fixed $R'$ values because the calculated $\tau'$ is almost linear in $d'^2$. 

On the basis of Fig.~\ref{fig:extauKBvsR} and Fig.~\ref{fig:tauKBvsdsq}, we assume that the numerical value of $\tau'$ is a function of $R'$ and $d'$ of the form
\begin{equation}\label{eq:eq0}
\tau'(R',d') \sim \bar{\tau}' + c_1 d'^2 + c_2 R' d'^2 + h(d'^2,R'),
\end{equation}
where $c_1$ and $c_2$ are constants, $\bar{\tau}'$ is the line tension nearly invariant of the grid spacing $d'$ and distance $R'$, and $h(d'^2,R')$ represents terms of higher order than $d'^2$ and $R'$. From the expression \eqref{eq:eq0} of the numerical line tension $\tau'(R',d')$, we can eliminate approximately the numerical error which depends on $R'$ by linear extrapolation of $\tau'$ from various $R'$ toward $R'=0$ as in Fig.~\ref{fig:extauKBvsR} and obtain a correction of $\tau'$ at $R'=0$ , which is given by
\begin{equation}\label{eq:eq1}
\tau'_0(d') \equiv \tau'(R'=0,d') \sim  \bar{\tau}' + c_1 d'^2 + h_0(d'^2),
\end{equation}
where $h_0(d'^2)$ represents terms of higher order than $d'^2$. The extrapolated results of $\tau'_0$ \eqref{eq:eq1} are listed in Table~\ref{tab:taurefinement1}.
\begin{table}
\caption{\label{tab:taurefinement1}Refinement of line tension $\tau'$ based on the Kerins-Boiteux formula. $\tau'_0$ is the extrapolated line tension at $R'=0$ for various $d'$, where the dominant term is $d'^2$. $\tau'_1$ is the first level correction of $\tau'_0$ by eliminating the $d'^2$ term.}
\begin{ruledtabular}
\begin{tabular}{ccccc}
& $d'$=0.05 & $d'$=0.1 & $d'$=0.2 & $d'$=0.4 \\
\hline
$\tau'_0$ & -0.07216833 & -0.07216553 & -0.07215562 & -0.07211596 \\ 
$\tau'_1$ & -0.07216926 & -0.07216883 & -0.07216884 & N/A \\
\end{tabular}
\end{ruledtabular}
\end{table}

In addition, we can refine our result at $R'=0$ by using Richardson's extrapolation \cite{richardson1911approximate,richardson1927deferred}, in which results for two successive $d'$ values are used to eliminate the $d'^2$ term. The first level of correction is defined as
\begin{equation}\label{eq:eq2}
\tau'_1(d') = \frac{4\tau'_0(d')-\tau'_0(2d')}{3} = \bar{\tau}' + h_1(d'^2),
\end{equation}
where $h_1(d'^2)$ represents the terms of higher order than $d'^2$. From the calculation of $\tau'_1$ in Table~\ref{tab:taurefinement1}, the line tension $\tau'$ for $a=2/3$ is approximated by
\begin{equation}
\bar{\tau}' \sim-0.072169,
\end{equation}
where the uncertainty is in the final digit. It is well known both theoretically \cite{rowlinson2002molecular,widom2006models} and experimentally \cite{adamson1997physical,amirfazli2004status} that line tensions, unlike interfacial tensions, can be either positive or negative. Physically, a negative line tension means, for example, that the line of intersection of a sessile drop with a substrate would tend to expand \cite{adamson1997physical,widom1995line}, but is ultimately limited by positive interfacial tensions.


\subsection{Scaling of the Density Functional Model for Line Tension}

In our original way of scaling, we factored out $B \ell^2$ from the excess grand potential and also from the Kerins-Boiteux formula for line tension. Then, we calculated the integral in a dimensionless domain. However, our potential is parametrized by $a$, which means that solutions of the Euler-Lagrange equations and the calculation of $\tau'$ depend on $a$. Here, to elucidate the $a$-dependence of our model, we study the problem in a new framework by defining the following new scaled variables,
\begin{equation}
Y_i \equiv \frac{X_i-b}{a-b}=\frac{2 X_i-(1-a)}{3 a-1},
\end{equation}
where $\sum_{i=1}^3 Y_i = 1$ and $a\neq 1/3$. In our model, the value of $X_i$ is limited from $b$ to $a$, so $Y_i $ varies from 0 to 1. In terms of the new scaling variables, the new dimensionless form of the excess grand potential \eqref{eq:omegaxsgeneral} for a symmetric three-phase contact line is given by
\begin{equation}
\tilde{\Omega}_{xs} \equiv \frac{\Omega_{xs}}{\tilde{B} L \tilde{\ell}^2} = \int_A (\tilde{f}(Y_1,Y_2) + \tilde{g}(\tilde{\nabla} Y_1,\tilde{\nabla} Y_2)) \mathrm{d}\tilde{A},
\end{equation}
where we define the following new scaled constants $\tilde{B} \equiv B (a-b)^4$, $\tilde{\ell}^2 \equiv \ell^2/(a-b)^2$, $\tilde{\nabla} \equiv \tilde{\ell}\nabla$, $\tilde{A} \equiv A/\ell^2$. The scaled potential and gradient energy density are
\begin{equation}\label{eq:scaledf}
\tilde{f}(Y_1,Y_2,Y_3) = \sum\limits_{i=1}^3 (Y_i)^2(Y_i-1)^2
\end{equation}
and
\begin{equation}\label{eq:scaledg}
\tilde{g}(\tilde{\nabla} Y_1,\tilde{\nabla} Y_2,\tilde{\nabla} Y_3)  = \sum\limits_{i=1}^3 \lvert \tilde{\nabla} Y_i \rvert^2.
\end{equation}
In terms of independent variables, we define the two-variable functions
\begin{equation}
\tilde{f}(Y_1,Y_2) \equiv \tilde{f}(Y_1,Y_2,1-Y_1-Y_2)
\end{equation}
and
\begin{equation}
\tilde{g}(\tilde{\nabla} Y_1,\tilde{\nabla} Y_2) \equiv \tilde{g}(\tilde{\nabla} Y_1,\tilde{\nabla} Y_2,-\tilde{\nabla} Y_1-\tilde{\nabla} Y_2)
\end{equation}
Similarly, we obtain a new expression of the dimensionless line tension \eqref{eq:dimensionlessKB} in the form of the Kerins-Boiteux formula \cite{KBpaper}:
\begin{equation}\label{eq:tildetau}
\tilde{\tau} \equiv  \frac{\tau}{\tilde{B} \tilde{\ell}^2} = \int_A (-\tilde{f}(Y_1,Y_2) + \tilde{g}(\tilde{\nabla} Y_1,\tilde{\nabla} Y_2)) \mathrm{d}\tilde{A}.
\end{equation}
Note that the integral of $\tilde{\Omega}_{xs}$ and $\tilde{\tau}$ are both independent of $a$ and dimensionless. 

Based on the numerical methods presented in Sec.~\ref{sec:numerical}, we can compute $\tilde{\tau}$ and refine the result by Richardson's extrapolation as in Table~\ref{tab:tautilderefinement}. The refined $\tilde{\tau}$ is
\begin{equation}\label{eq:tildetauvalue}
\tilde{\tau} \sim -0.28868,
\end{equation}
where the uncertainty is in the last digit.
\begin{table}[h!]
\caption{\label{tab:tautilderefinement}Refinement of the scaling line tension $\tilde{\tau}$ based on Kerins-Boiteux formula. $\tilde{\tau}_0$ is the extrapolated line tension at $\tilde{R}=0$ for various $\tilde{d}$ values, where the dominant term is $\tilde{d}^2$. $\tilde{\tau}_1$ is the first level correction of $\tilde{\tau}_0$ by eliminating the $\tilde{d}^2$ term.}
\begin{ruledtabular}
\begin{tabular}{ccccc}
& $\tilde{d}=0.05$ & $\tilde{d}=0.1$ & $\tilde{d}=0.2$ & $\tilde{d}=0.4$ \\
\hline
$\tilde{\tau}_0$ & -0.28866211 & -0.28862233 & -0.28846449 & -0.28781290 \\ 
$\tilde{\tau}_1$ & -0.28867521 & -0.28867560 & -0.28868169 & N/A \\
\end{tabular}
\end{ruledtabular}
\end{table}
%

\begin{figure}
\includegraphics[scale=0.89]{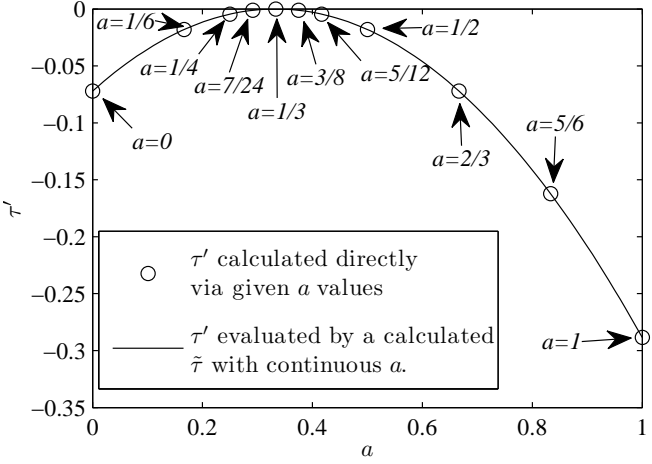}
\caption{$\tau'$ as a function of $a$. The little circles indicate the refined $\tau'$ results calculated directly from various $a$ values. The curve is a plot of $\tau'$ based on a refined calculation of $\tilde{\tau}$ as shown in Table~\ref{tab:tautilderefinement}.}
\label{fig:tau-a}
\end{figure}

According to the new scaled constants, we find that the dimensionless line tension $\tau'$ \eqref{eq:dimensionlessKB} and its new scaled expression $\tilde{\tau}$ \eqref{eq:tildetau} obey the following relation
\begin{equation}\label{eq:tauprimeadependence}
\tau' =\frac{\tau}{B\ell^2}= \left(\frac{3}{2}\right)^2 \left(a-\frac{1}{3}\right)^2 \tilde{\tau}
\end{equation}
which shows that $\tau'$ is proportional to $(a-1/3)^2$ as indicated in Fig.~\ref{fig:tau-a}, and $\tau'$ is equal to $\tilde{\tau}$ for $a=1$. In Fig.~\ref{fig:tau-a}, we use the numerical value of $\tilde{\tau}$ \eqref{eq:tildetauvalue} and the relation \eqref{eq:tauprimeadependence} that connects $\tau'$ and $\tilde{\tau}$ to plot a curve in Fig.~\ref{fig:tau-a}, which agrees with the numerical values of $\tau'$ for various values of $a$ in the same figure. These numerical values of $\tau'$ were obtained by the same numerical methods presented in Sec.~\ref{sec:numerical} and refined by Richardson's extrapolation.

Because the temperature-dependent parameter $a$ is proportional to a density, it should approach its critical value $a_c=1/3$, as $\lvert T-T_c\rvert^{1/2}$, according to the predictions of mean-field theory \cite[p.~251]{rowlinson2002molecular}. Because surface tension vanishes as $\lvert T-T_c\rvert^{3/2}$ in mean-field theory, this explains the factor $\lvert a-1/3 \rvert^3$ in \eqref{eq:gamma12} and \eqref{eq:symsigma}. Moreover, the results of Varea and Robledo \cite{varea1992statistical} in the mean-field approximation show that the ratio of critical exponents of line tension and surface tension is $2/3$, consistent with the ratio of $\tau'$ in \eqref{eq:tauprimeadependence} and $\sigma'$ in \eqref{eq:symsigma}, namely
\begin{equation}
\frac{\tau'}{\sigma'} \propto \frac{\lvert a-\frac{1}{3}\rvert^2}{\lvert a-\frac{1}{3}\rvert^3}.
\end{equation}
The authors are grateful to one of the reviewers for pointing out this observation. Thus, when the system approaches a homogeneous solution, the line tension vanishes more slowly than the interfacial tension.

\begin{figure}
\subfigure[\;$a_i \geq b_i$]{
\label{subfig:gibbsasym}
\includegraphics[scale=0.185]{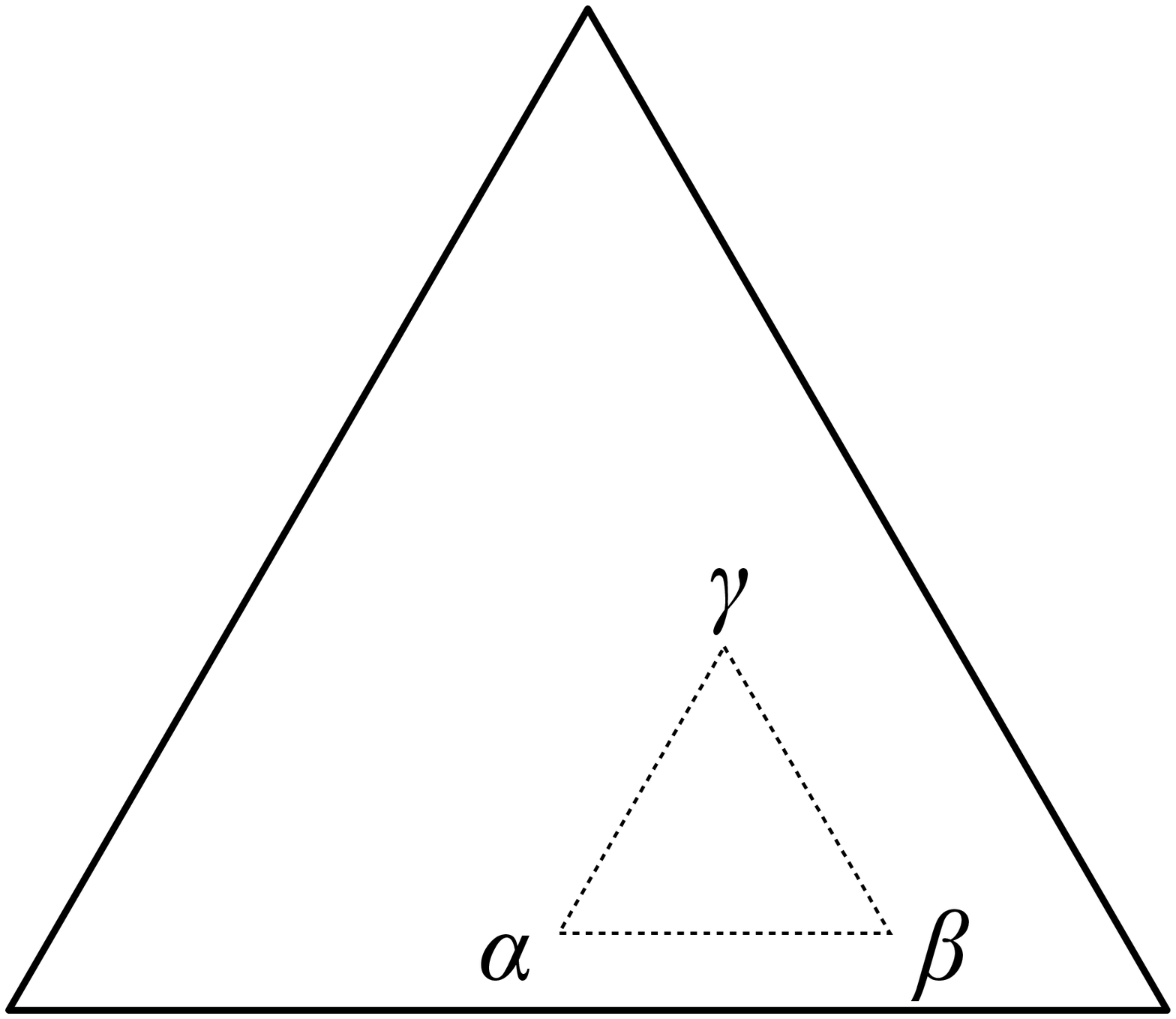}}
\subfigure[\;$a_i \leq b_i$]{
\label{subfig:gibbsasyminverted}
\includegraphics[scale=0.185]{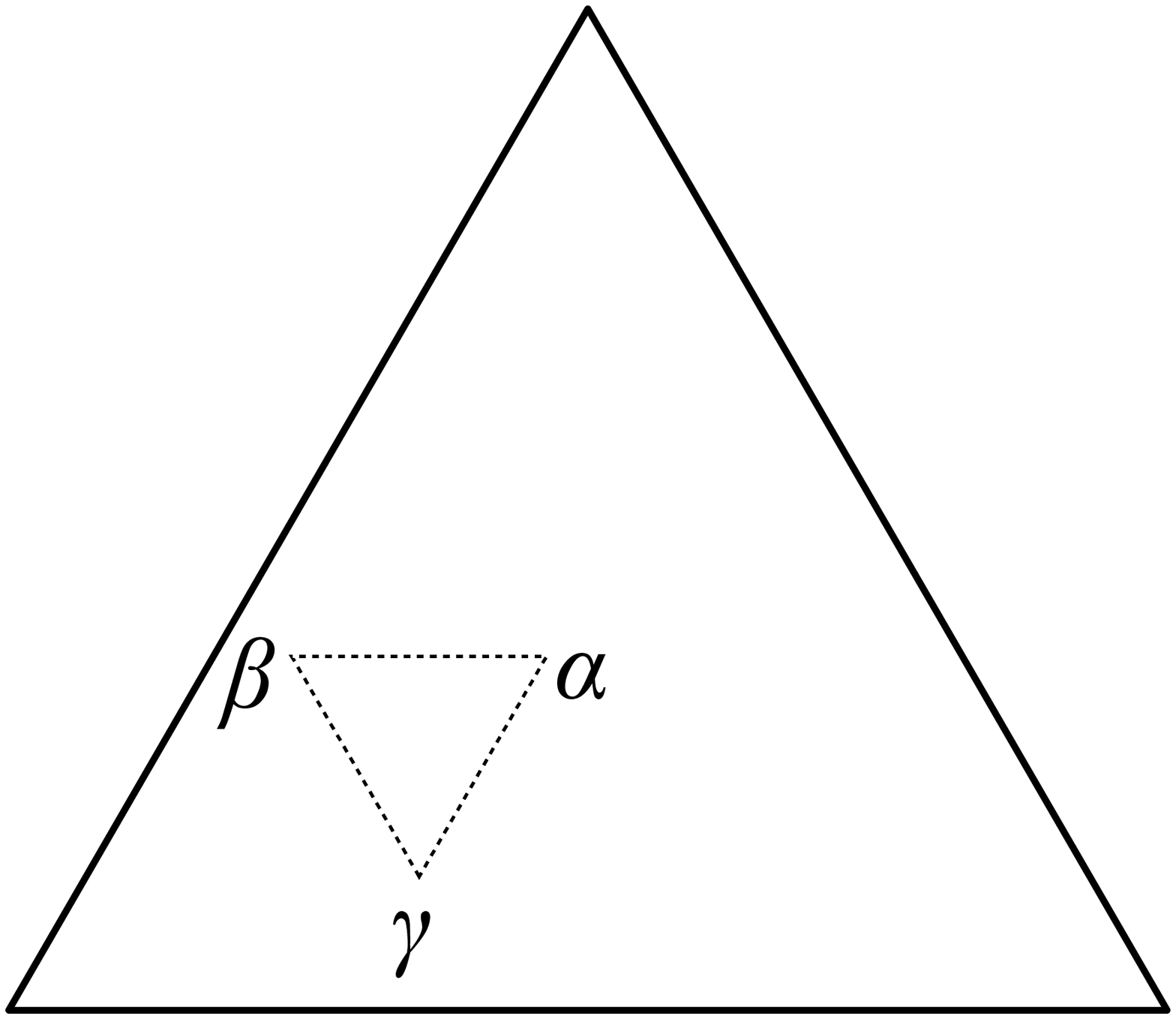}}
\caption{\label{fig:gibbscompare} Examples of the location of bulk phases $(\alpha,\beta,\gamma)$ for the potential $f^{\ast}$. \subref{subfig:gibbsasym} For allowed $a_i$ and $1\leq Q\leq 3$, we have $a_i \geq b_i$ and the phases are at the vertices of equilateral triangles that are magnifications of the Gibbs triangle. \subref{subfig:gibbsasyminverted} For $0\leq Q \leq 1$, the bulk phases are at the vertices of equilateral triangles that are inverted with respect to the Gibbs triangle. If $Q = 1$, we have $a_i = b_i$ and the triangles degenerate to points (bulk criticality).}
\end{figure}

We note that a somewhat more general potential, namely 
\begin{equation}\label{eq:asymfinx}
f^{\ast}(X_1,X_2,X_3) = \sum_{i = 1}^3 (X_i-a_i)^2(X_i-b_i)^2,
\end{equation}
containing the six constants $0\leq a_i \leq 1$ and $0\leq b_i \leq 1$, can be mapped onto the potential $\tilde{f}$ in \eqref{eq:scaledf}.  In this case, the minima are located at the bulk phases $\alpha = (a_1,b_2,b_3)$,  $\beta = (b_1,a_2,b_3)$,  and $\gamma = (b_1,b_2,a_3)$.  The condition $\sum_{i =1}^3X_i = 1$ leads to the three constraints $a_1+b_2+b_3 = 1$,  $b_1+a_2+b_3 = 1$ and $b_1+b_2+a_3 = 1$.  Regarding the $a_i$ to be independent variables,
\begin{equation}\label{eq:birelation}
b_i = (1/2)[1 + a_i - a_j - a_k] = (1-Q)/2 + a_i,
\end{equation}
where $i$, $j$ and $k$ are all different and $Q = \sum_{\ell=1}^{3} a_\ell$. Since $0\leq Q \leq 3$, we have $b_i\geq a_i$ for $0\leq Q \leq 1$ and $b_i\leq a_i$ for $1\leq Q \leq 3$.  Any choice of the vector $(a_1,a_2,a_3)$ in the positive unit cube will lead to $b_i \leq{1}$ but the requirement  $0\leq b_i$ restricts $(a_1,a_2,a_3)$  to lie within the positive unit cube truncated by a pyramid consisting of three planes; the apex of the pyramid is located at $(1,1,1)$ and the other three vertices are located at $(1,0,0)$, $(0,1,0)$  and $(0,0,1)$. This truncation only restricts
$(a_1,a_2,a_3)$ if  $1\leq Q \leq 3$. It turns out that the three phases $\alpha,\beta,\gamma$ are located at the vertices of equilateral triangles that lie within or on the Gibbs triangle and whose sides are parallel to the sides of the Gibbs triangle, as depicted in Figure \ref{fig:gibbscompare}.  For allowed 
$(a_1,a_2,a_3)$ and $1\leq Q \leq 3$, the phases $\alpha,\beta,\gamma$ are located at the vertices of triangles that are magnifications of the Gibbs triangle, as depicted in Figure \ref{subfig:gibbsasym}.  For $0\leq Q \leq 1$, the phases lie at the vertices of equilateral triangles that are inverted with respect to the Gibbs triangle, as depicted in Figure \ref{subfig:gibbsasyminverted}. 

By defining the new variables $Z_i =2(X_i - a_i)/(1-Q)$, which satisfy $\sum_{i=1}^3 Z_i =1$, the potential $f^{\ast}$ becomes 
\begin{equation}\label{eq:asymfinz}
f^{\ast} = \left(\frac{1-Q}{2}\right)^4\sum_{i=1}^3 Z_i^2(1-Z_i)^2,
\end{equation}
which has the same form as $\tilde{f}$ in \eqref{eq:scaledf}. Thus, the potential $f^{\ast}$ is actually a shifted and scaled version of the potential $\tilde{f}$, resulting in $\tau' = \tau/(B\ell^2) = [(1-Q)/2]^2\tilde{\tau}$. The phases merge (bulk criticality) whenever $Q = 1$. 

\section{Line Adsorption}


The Gibbs adsorption equation relates the change of interfacial tension to the change of field variables, the coefficients being surface adsorptions. In our case, the Gibbs adsorption equation for the $\alpha \beta$-interface is
\begin{equation}\label{eq:gibbsadsorption}
\mathrm{d}\sigma_{\alpha \beta}=- \left(\Gamma_1^{\alpha \beta} \mathrm{d}M_1 + \Gamma_2^{\alpha \beta} \mathrm{d}M_2 +   \Gamma_T^{\alpha \beta} \mathrm{d}T \right),
\end{equation}
where $\Gamma_i$ is the adsorption (surface excess per unit area) of the chemical constituent $i$, and $\Gamma_T$ is the adsorption related to entropy. Notice that each of the surface adsorptions, $\Gamma_1^{\alpha \beta}$, $\Gamma_2^{\alpha \beta}$, and $\Gamma_T^{\alpha \beta}$, depends on the choice of dividing surface, but $d\sigma_{\alpha \beta}$ in the Gibbs adsorption equation \eqref{eq:gibbsadsorption} is independent of this choice. Independence of location of the dividing surface is one of the properties of the Gibbs adsorption equation. A similar relation also works for the $\beta \gamma$- and $\gamma \alpha$-interfaces.

The Gibbs adsorption equation for diffuse interface models have been studied extensively (See Rowlinson and Widom \cite[p~37-38]{rowlinson2002molecular}). As an extension of the Gibbs adsorption equation, Djikaev and Widom \cite{djikaev2004geometric} introduce a line adsorption equation, which depends on the choice of the position of the contact line $\vec{r}$. The line adsorption equation of Djikaev and Widom is
\begin{equation}\label{eq:dtau0}
\begin{split}
\mathrm{d}\tau = & -\sum\limits_{i=1}^{c+1} \Lambda_i(\vec{r}) \mathrm{d}\mu_i \\
&- \left(\vec{e}_{\alpha\beta} \mathrm{d}\sigma_{\alpha\beta} + \vec{e}_{\beta\gamma} \mathrm{d}\sigma_{\beta\gamma} + \vec{e_{\gamma\alpha}} \mathrm{d}\sigma_{\gamma\alpha} \right) \cdot \left( \vec{r} - \vec{r}_0 \right),
\end{split}
\end{equation}\label{eq:dtauinv}
where $\Lambda_{c+1}$ is the line adsorption corresponding to the entropy conjugate to $T$. This shows that an infinitesimal change of the line tension $d\tau$, for a $c$-component system, comes from two parts. The first part is analogous to the Gibbs adsorption equation, which is a linear combination of the infinitesimal changes of the field variables $\mu_i$ multiplied by the line adsorption $\Lambda_i$, that depends on the position of the three-phase contact line. The second part is the inner product of the difference between $\vec{r}$ and $\vec{r}_0$, a specific choice of $\vec{r}$, and the summation of the infinitesimal changes of the three interfacial tensions $d\sigma_{k}$ multiplied by the corresponding unit vector $\vec{e}_k$ along the interface $k$ and perpendicular to the contact line. For a given value of $\vec{r}_0$, the line adsorption equation \eqref{eq:dtau0} does not depend on $\vec{r}$. The second term arises because the interfaces can change angles as the $\mu_i$ change. As shown by \cite{djikaev2004geometric,taylor2005adsorption,koga2006line}, the second term can be eliminated if $\vec{r_0}$ is chosen to lie along a special line. In that case, the line adsorption equation \eqref{eq:dtau0} becomes
\begin{equation}
\mathrm{d}\tau = -\sum\limits_{i=1}^{c+1} \Lambda_i(\vec{r})\mathrm{d}\mu_i.
\end{equation}

In our case of a symmetric three-phase contact line, the second part of the line adsorption equation \eqref{eq:dtau0} is zero, since the three interfacial tensions remain equal as $a$ changes (See the form \eqref{eq:symsigma} of surface tension) and the summation of the three unit vectors is zero. For our case, the line adsorption equation \eqref{eq:dtau0} becomes
\begin{equation}\label{eq:dtau1}
\mathrm{d}\tau=- \left(\Lambda_1(\vec{r})\mathrm{d}M_1 +  \Lambda_2(\vec{r})\mathrm{d}M_2 + \Lambda_T(\vec{r})\mathrm{d}T\right) ,
\end{equation}
where $\Lambda_i$ is the line adsorption corresponding to chemical constituent $i$ and $\Lambda_T$ is the line adsorption corresponding to the entropy. From the form \eqref{eq:dtau1} of the line adsorption equation, it appears that $\tau$ depends on three variables $M_1$, $M_2$, and $T$. However, if we consider the two Clapeyron equations for this three-phase system,
\begin{equation}\label{eq:clapeyron1}
\begin{split}
(\rho_1^{\alpha}-\rho_1^{\beta})\mathrm{d}M_1 + (\rho_2^{\alpha}-\rho_2^{\beta})\mathrm{d}M_2 + (s^{\alpha}-s^{\beta})\mathrm{d}T & =0 \\
(\rho_1^{\beta}-\rho_1^{\gamma})\mathrm{d}M_1 + (\rho_2^{\beta}-\rho_2^{\gamma})\mathrm{d}M_2 + (s^{\beta}-s^{\gamma})\mathrm{d}T & =0,
\end{split}
\end{equation}
there is only one independent variable for $\tau$, which could be $M_1$, $M_2$, or $T$. For instance, if $d\tau$ only depends on $T$, we have
\begin{equation}\label{eq:dtaudt}
\mathrm{d}\tau=-\Lambda_T^{eff} \mathrm{d}T,
\end{equation}
where $\Lambda_T^{eff}$ is a linear combination of all $\Lambda_i$ and is invariant. By locating our contact line at the center of our triangular domain, the symmetry of our potential leads to $\Lambda_1=\Lambda_2=0$, so $\Lambda_T^{eff}=\Lambda_T$. A similar simplification also applies for the Gibbs adsorption equation \eqref{eq:gibbsadsorption}. From the Clapeyron equations \eqref{eq:clapeyron1}, $\mathrm{d}M_1$, $\mathrm{d}M_2$, and $\mathrm{d}T$ are linearly related and since (with $B\ell^2=$constant) there is only one variable $a$ in the problem, we can write
\begin{equation}\label{eq:dtauda}
\mathrm{d}\tau=-\Lambda_a^{eff} \mathrm{d}a,
\end{equation}
where $\Lambda_a^{eff}=\Lambda_T \mathrm{d}T/\mathrm{d}a$ is an effective line adsorption corresponding to $a$. From the relation of $\tau'$ and $\tilde{\tau}$ \eqref{eq:tauprimeadependence}, we calculate
\begin{equation}\label{eq:lambdaa}
\Lambda_a^{eff}=\frac{\mathrm{d}\tau}{\mathrm{d}a}=2B\ell^2 \left(\frac{3}{2}\right)^2\left( a-\frac{1}{3}\right) \tilde{\tau}
\end{equation}

\section{Summary and Conclusions}

A three-phase contact line in a three-phase fluid system is studied by a mean-field density functional model, in which classical sharp fluid-fluid interfaces are replaced by diffuse interfaces. The geometry of the system is chosen to be a prism, where each of its lateral faces is perpendicular to one of the interfaces and both the cap and bottom are Neumann triangles. To define a tractable model, we assume that the intermolecular forces are short range and can be modeled by local densities. The dimension of the system is large compared to the interfacial width. The excess grand potential of the system is modeled by a functional consisting of a highly symmetric three-well potential and a gradient energy, which is linear in the squared gradients of the three compositions (in terms of mole fractions). We assume for simplicity that the molar volume is a constant, so there are only two independent densities. We use a variational approach to find the governing coupled Euler-Lagrange equations. In the far-field limit, where the distance from the contact line is large compared to the interfacial width, the transition between two bulk phases having different chemical compositions is essentially one-dimensional. Analytically, a far-field asymptotic solution is obtained and is used to calculate the interfacial tensions. This connects our phenomenological model to interfacial tensions and to the equilibrium angles for classical sharp interfaces. 

Because of the nonlinearity of our Euler-Lagrange equations, we cannot find a near-field asymptotic solution. Instead, we perform a numerical analysis for a symmetric three-phase contact line. By applying a triangular grid that fills the entire domain, we implement a consistent discretization to obtain the discrete Euler-Lagrange Equations from the variation of a discretized excess grand potential. To solve the system of these coupled algebraic equations for the entire domain, we apply a successive over-relaxation method and use the asymptotic far-field solutions as the boundary conditions. The calculated isoconcentrates (constant mole fractions) bend and the effective interfacial width increases slightly near the contact line. Close to the outer boundary, the nearly parallel isoconcentrates along the diffuse interfaces show that our domain size is close to the asymptotic regime, so the boundary conditions are sufficient.

We study the line tension associated with a symmetric three-phase contact line based on our mean-field density functional model, which is the excess grand potential over the entire domain diminished by the energies of the surfaces extrapolated from the interfacial tensions in the far-field. By using the Kerins-Boiteux formula, which formulates the expression of the line tension into a single integral, we calculate the line tension and analyze the corresponding integrand. Our results show that the numerical values of the line tension require a correction proportional to the domain size and to the square of the grid spacing. To refine our result, we eliminate approximately the error associated with the domain size by linear extrapolation of the values of line tension from finite sizes to zero. Furthermore, we use Richardson's method to reduce the error associated with the square of the grid spacing to obtain the next level of refinement. The calculation of line tension based on our mean-field density model shows that the value of line tension is negative and proportional $(a-1/3)^2$, where $a$ is a parameter in our model. We introduce a scaling method to resolve this relation for our model. The line tension is proportional to $(a-1/3)^2$ multiplied by an integral (negative and independent of $a$), in agreement with independent calculations for various values of $a$. In contrast, the far-field interfacial tension is proportional to $(a-1/3)^3$. When $a=1/3$, both the line tension and interfacial tension vanish. Physically, this means that the three chemical constituents share the same value of mole fraction ($1/3$) and are equally and uniformly distributed over the entire system, a single phase. In effect, the interfacial width, which is reciprocal to $(a-1/3)$, is infinite. On the other hand, we can either say that the contact line and three interfaces vanish or occupy the entire domain. However, when $a$ approaches $1/3$, the interfacial tensions decay faster than the line tension.

Finally, we relate the change of line tension to the line adsorptions by \cite{djikaev2004geometric}. Thermodynamically, we show that there is only one independent field which could be chosen as temperature. We are able to link it to the line adsorption corresponding to $a$, since $a$ is the only variable in our model (if other coefficients $B$ and $\ell$ are treated as constants). Consequently, we find an analytical expression of the line adsorption corresponding to $a$.

In order to link our model to realistic systems, we make the following numerical estimates. For $T \sim 300~\mathrm{K}$, typical values of interfacial tension are a few times of $10^{-2}~\mathrm{N/m}$, and interfacial widths are a few \AA~\cite{rowlinson2002molecular}. We assume $a=1$, the interfacial tension $\sigma \sim 5\times 10^{-2}~\mathrm{N/m}$, and the characteristic length $\ell \sim 1$ \AA, corresponding to an interfacial width of $2\sim 3$~\AA. Inserting the form \eqref{eq:symsigma} for interfacial tension into the form \eqref{eq:tauprimeadependence} for line tension yields
\begin{equation}
\frac{\tau}{\sigma}=\frac{\sqrt{2}}{\lvert a-\frac{1}{3}\rvert} \tilde{\tau} \ell.
\end{equation}
We obtain $\tau \sim -0.3 \times 10^{-11}~\mathrm{N}$. The magnitude of $\tau$ is at the lower end of typical experimental values, which are in the range $10^{-11}$ to $10^{-9}~\mathrm{N}$ \cite{amirfazli2004status}. $\Lambda_a^{eff} \sim -0.91 \times 10^{-11}~\mathrm{N}$. A crude estimate gives $\Lambda_T \sim - \tau/T \sim 10^{-14}~\mathrm{N/K}$. The units of $\Lambda_T$ are entropy per unit length.

\begin{acknowledgments}
We appreciate the support of Department of Physics, Carnegie Mellon University for this work. Thanks are due to Christopher F. Eldred for introducing the symmetric potential $f$ and using it to calculate compositions and interfacial tensions far from the three-phase contact line. We thank Benjamin Widom for extensive theoretical discussions of line tensions and adsorptions as well as a critical reading of our manuscript. We also thank Steve Garoff for his intuitive ideas from an experimental aspect and Shlomo Ta'asan for valuable advice on numerical methods.
\end{acknowledgments}

\bibliography{capillarity}
\bibliographystyle{apsrev4-1}

\end{document}